\documentclass[twocolumn,preprintnumbers,superscriptaddress,landscape,nofootinbib]{revtex4}
\usepackage[utf8]{inputenc}
\usepackage{amsmath}
\usepackage{amsthm}
\usepackage{amssymb}
\usepackage{float}
\usepackage{braket}
\usepackage{graphicx}
\usepackage{url}
\usepackage{here}
\usepackage{natbib}
\usepackage{rotating}
\usepackage[colorlinks=true, pdfstartview=FitV, linkcolor=red, citecolor=blue, urlcolor=blue]{hyperref}
\usepackage{slashed}
\usepackage{multirow}
\usepackage[normalem]{ulem}
\graphicspath{{./figures/}}

\newcommand{\be}{\begin{equation}}      
\newcommand{\ee}{\end{equation}}      
\newcommand{\bea}{\begin{eqnarray}}      
\newcommand{\eea}{\end{eqnarray}}

\newcommand{\Tr}{\mathrm{Tr}}

\newcommand{\ctext}[1]{\raise0.2ex\hbox{\textcircled{\scriptsize{#1}}}}

\usepackage{tikz}
\usetikzlibrary{quantikz}

\theoremstyle{definition}

\theoremstyle{remark}

\usepackage[normalem]{ulem}

\begin{document}

\title{Criticality of quantum energy teleportation at phase transition points in quantum field theory} 
\author{Kazuki Ikeda}
\email[]{kazuki7131@gmail.com}
\affiliation{Co-design Center for Quantum Advantage, Department of Physics and Astronomy, Stony Brook University, Stony Brook, New York 11794-3800, USA}
\affiliation{Center for Nuclear Theory, Department of Physics and Astronomy, Stony Brook University, Stony Brook, New York 11794-3800, USA}

\bibliographystyle{unsrt}

%%%%%%%%%%%%%%%%%%%%%%%%%%%%%%%%%%%%%%
\begin{abstract}
Quantum field theory can be a new medium for communication through quantum energy teleportation. We performed a demonstration of quantum energy teleportation with a relativistic fermionic field theory of self-coupled fermions, called the massive Thirring model. Our results reveal that there is a close relation between the amount of energy teleported and the phase diagram of the theory. In particular, it is shown that the teleported energy peaks near the phase transition points. The results provide new implications for phase diagrams of field theory in terms of quantum communication and quantum computing.
\end{abstract}

\maketitle

%%%%%%%%%%%%%%%%%%%%%%%%%%%%%%%%%%%%%%
\section{Introduction}
Quantum field theory (QFT) has been quite successful in explaining quantum many-body systems. From condensed matter physics, such as superconductors and topological insulators, to the Standard Model of elementary particles as a low-energy effective theory of high-energy physics, QFT can explain a wide variety of experimental results with extremely high precision. The approach to non-perturbative phenomena is a remaining challenge for QFT, which has been explored by various methods such as first-principles calculations and lattice QCD. In addition, with the advent of quantum computers, we are able to perform real-time non-perturbative quantum simulations of many-body systems. One of the key challenges in studying QFT is the complexity of the calculations involved. Simulating these systems using classical computers can be computationally expensive, as the complexity of the calculations increases rapidly with the size of the system. Quantum computers, on the other hand, have the potential to perform these simulations much more efficiently. In addition to this, the development of quantum algorithms and quantum computers has greatly contributed to the fundamental understanding of quantum mechanics, including the control of quantum states and the measurement of quantum states.

\begin{figure*}
    \centering
    \includegraphics[width=\linewidth]{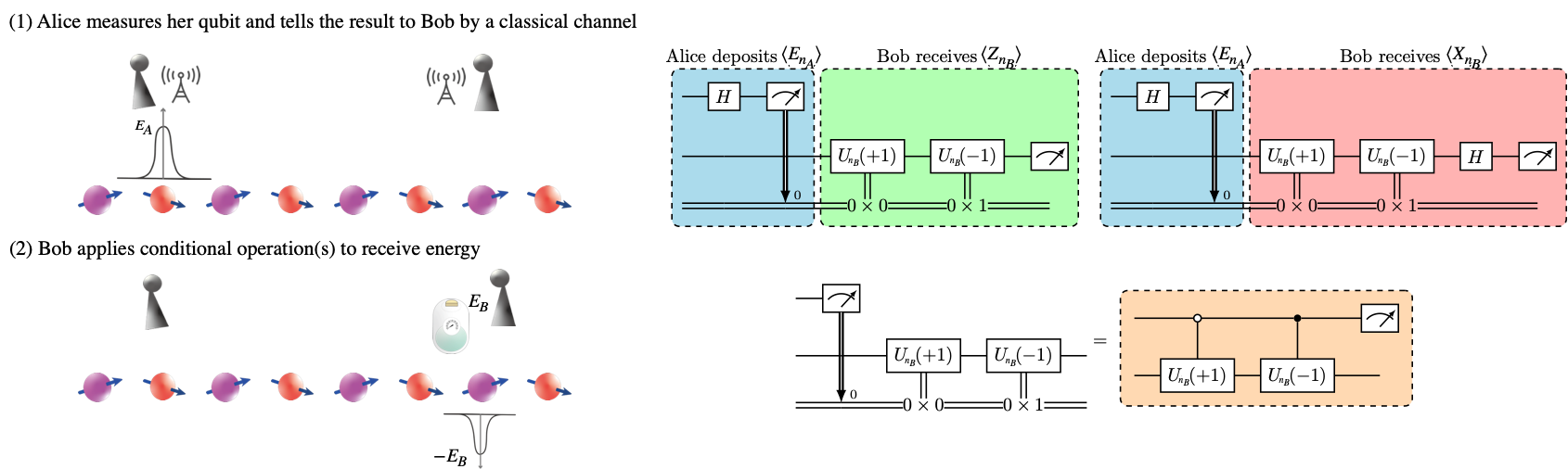}
    \caption{Protocol of quantum energy teleportation [Left] and the corresponding quantum circuits [Right]. First, Alice measures her local operator $X_{n_A}$ and tells her result ($\mu\in\{+1,-1\}$) to Bob. At this point, Alice's local energy is excited $E_{n_A}>0$. Then, to obtain energy, Bob applies conditional operation $U_{n_B}(\mu)$ to his local qubit and measures the corresponding terms of his local Hamiltonian $H_{n_B}$. Statistically he will observe $\langle H_{n_B}\rangle=\Tr[\rho_\text{QET}H_{n_B}]<0$ and gain $E_{n_B}=-\langle H_{n_B}\rangle$ through his measurement device.}
    \label{fig:protocol}
\end{figure*}

As such, understanding the behavior of quantum many-body systems through quantum simulations has been the primary focus of recent cross-disciplinary interest in physics and computer science, but for physics, the connection to quantum science and technology is not limited to quantum computation. Regarding the connection between QFT and quantum information theory, there are active studies on entanglement entropy and black holes~\cite{2006PhRvL..96r1602R,2019JHEP...10..132Y,2007JHEP...09..120H}. These studies are mainly concerned with high-energy physics at the Planck scale. While such attempts have been extremely successful, new efforts to reveal the nature of quantum systems and spacetime through measurement have been active in recent years in a wide range of fields, including high-energy physics, condensed matter physics and quantum computation~\cite{PhysRevLett.128.010603,PhysRevResearch.4.013174,briegel2009measurement,2020PNAS..117.5706G,stephen2022universal,2022arXiv220304338K,2023arXiv230108775T,2022arXiv221005692M,PhysRevD.105.066011,2022arXiv220316269R,PhysRevD.105.065003}.

Quantum energy teleportation (QET) is a protocol for the study of local energies that takes advantage of the entanglement nature of the ground state of quantum many-body systems
~\cite{HOTTA20085671,2009JPSJ...78c4001H,2015JPhA...48q5302T,2009PhRvA..80d2323H,PhysRevA.82.042329,Hotta_2010,2023arXiv230102666I}.
Just as quantum teleportation can transfer quantum states to remote locations~\cite{PhysRevLett.70.1895,furusawa1998unconditional,2015NaPho...9..641P,takeda2013deterministic,thotakura2022quantum}, it is expected that QET can transfer energy to remote locations using local operation and classical communication (LOCC) only. The role of QET in physics and information engineering is largely unexplored, as the theory has not received much attention for long time since it was proposed about 15 years ago. An interesting property of QET is that multiple people in different locations, who share the same ground state initially, can simultaneously lower the energy of their local systems by applying conditional operations. This is only possible when the sender and receivers of the energy conduct the appropriate LOCC, and cannot be obtained by any unitary operation or random conditional operations.
Therefore QET will not only help to enhance our understanding of fundamental issues in quantum statistical mechanics, condensed matter physics, and high-energy physics but will also provide interesting perspectives for engineering applications of quantum computation and quantum communication. 

The purpose of this paper is to investigate the role and properties of QET in field theory. From the viewpoint of quantum computer applications, we simulate QET using the massive Thirring model (low dimensional quantum electrodynamics (QED)), which is one of the most widely used (1+1) dimensional models of QFT. First, we estimate the phase diagram of the massive Thrring model using entanglement entropy and chiral condensate. The main result of this paper is the identification of a sharp peak in teleported energy near the phase transition point. We also analyze the time-evolution of the entanglement entropy difference $\Delta S_{AB}$ using Alice's post-measurement state and show numerically how the entanglement entropy lost in Alice's measurement is recovered over time due to particle-particle interactions in the system if Bob does nothing after Alice's measurement. Some of the results in this paper are based on simulations of quantum gate operations using qasm\_simulator provided by IBM, and we confirm that all of these results are fully consistent with those obtained by exact diagonalization. These results provide new insights into local operations of quantum fields based on remote communication and non-trivial energy flow mediated by many-body quantum systems.

\section{Low dimensional QFT}
The (1+1) dimensional QFTs are of significant interest since they are simpler and more tractable than higher-dimensional QFT, and they have rich mathematical structures that have been studied extensively from various motivations, including condensed matter physics, high energy physics, statistical mechanics and mathematical physics~\cite{IZERGIN1982401}. Some of the models have a number of interesting properties, including confinement and the chiral anomaly therefore they are useful toy models of QCD. Typical models preferred in studies of (1+1) dimensional QFTs are the Thirring model and the Schwinger model. In particular, they are attractive models in terms of quantum simulation and quantum computation~\cite{PhysRevD.103.L071502,10.1093/ptep/ptac007,PhysRevResearch.2.023342,2020arXiv200100485C,Mishra_2020,klco2018quantum}.

The Thirring model is a simplified version of quantum electrodynamics (QED) in (1+1) dimensions, which was introduced by Walter Thirring in 1958~\cite{THIRRING195891}. It is a theory of a self–coupled Dirac
field, and it can be used to describe a variety of physical systems, such as superconductors~\cite{2001EPJC...20..723F}, statistical mechanics, high energy physics and mathematical physics~\cite{korepin1979direct}.

While the Thirring model and the Schwinger model are models for fermions, there is a significant (1+1) dimensional model for bosons, called sine-Gordon model, which is of significant interest in theoretical physics due to its integrability, soliton solutions, and relations to other models such as the Thirring model, massive Schwinger model, and to the $XY$-model. The sine-Gordon model is a (1+1) dimensional field theory that is described by the sine-Gordon equation, which is a nonlinear partial differential equation. The soliton solution of this model describes a kink or anti-kink solution which is a topological mode in the field that can be interpreted as particle like excitation~\cite{FADDEEV19781}. The topological nature of the solitons ensures the stability and the solitons retain their shape even during collision. 

It has been widely known that both models are related by the bosonization. By representing the fermionic fields in terms of bosonic fields, the bosonized version of the Thirring model becomes the sine-Gordon model. This is known as the $S$-duality between the two models. More detailed theoretical descriptions are given in Appendix~\ref{sec:quench}.

Throughout this work, we consider the massive Thirring model, whose Lagrangian is  
\begin{equation}
    \mathcal{L}_\text{Th}=\overline{\psi}(i\gamma^\mu\partial_\mu-m)\psi-\frac{g}{2}\overline{\psi}\gamma^\mu\psi\overline{\psi}\gamma_\mu\psi,
\end{equation}
where $m$ is the fermion mass, $g$ is the dimensionless four-fermion coupling constant and $\psi=\psi(x)$ is a spinor field with two components $\psi_1(x)$ and $\psi_2(x)$. It is widely known that the massive Thirring model is dual to the sine-Gordon model and the classical two-dimensional $XY$ model~\cite{PhysRevD.11.3026}. For example, a Kosterlitz-Thouless phase transition at $T\sim K\pi/2$ in the $XY$ model corresponds to a critical point $g\sim -\pi/2$, called Coleman’s instability point, in the Thirring model. They are also related with a critical point at $t\sim 8\pi$ in the sine-Gordon model. 

It turns out that the spin representation of the massive Thirring model is 
\begin{align}
\begin{aligned}
\label{eq:Ham_spin}
H_\text{Th}&=-\frac{1}{4a}\sum_{n=0}^{N-2}(X_nX_{n+1}+Y_nY_{n+1})\\
&+\frac{m}{2}\sum_{n=0}^{N}(-1)^{n+1}Z_n\\
&+\frac{\Delta(g)}{a}\sum_{n=0}^{N-2}\left(\frac{Z_n+1}{2}\right)\left(\frac{Z_{n+1}+1}{2}\right),
\end{aligned}
\end{align}
where $\Delta(g)=\cos\left(\frac{\pi-g}{2}\right)$, $a$ is the lattice spacing~\cite{PhysRevD.11.3026,PhysRevB.14.2153,PhysRevD.11.2088,2018slft.confE.229B,2019PhRvD.100i4504B,2017arXiv171009993B}. The theoretical background of the lattice Hamiltonian is described in Appendix~\ref{sec:Ham}.

\section{Simulation of Quantum Energy Teleportation}
\begin{figure*}
    \centering
    \includegraphics[width=\linewidth]{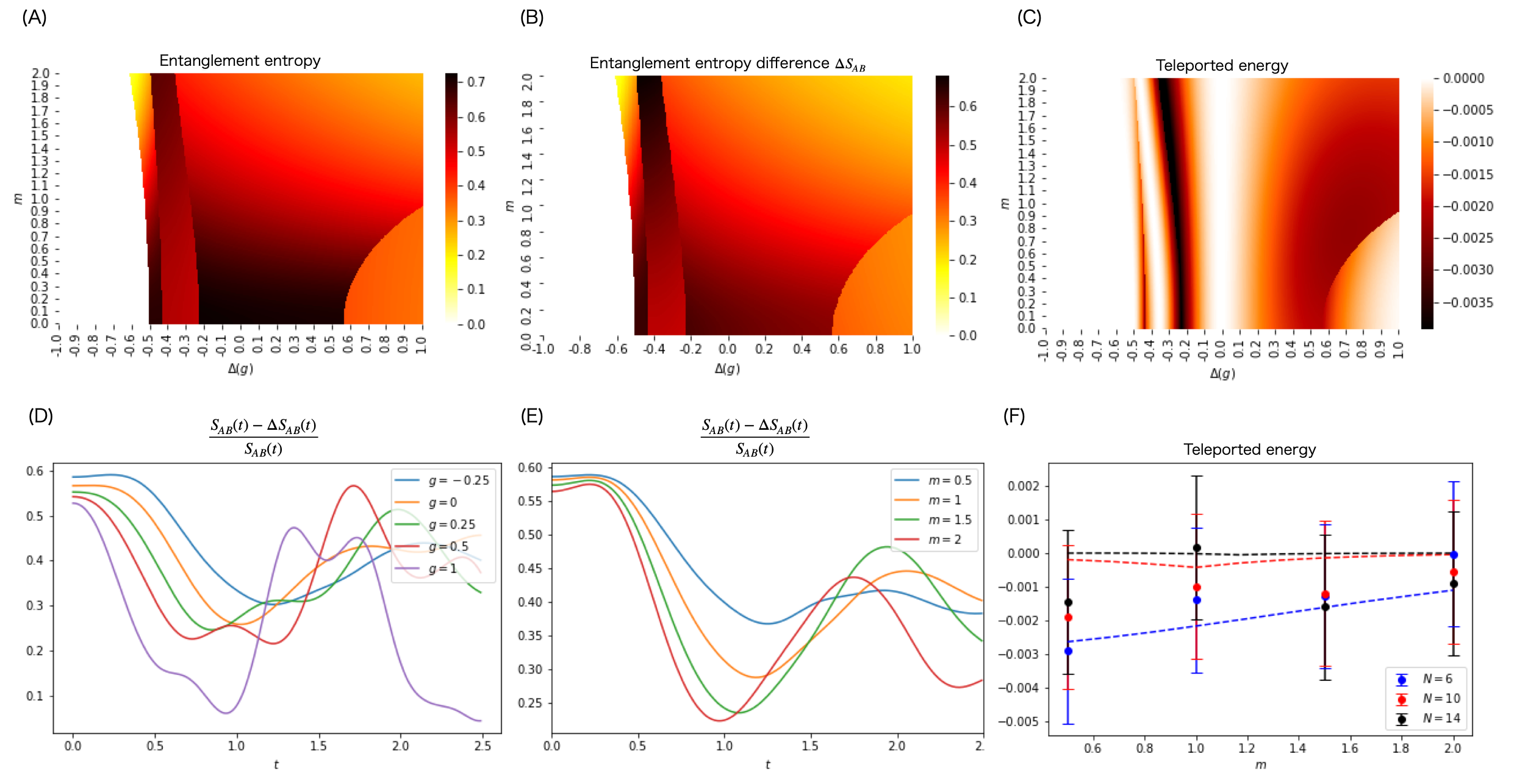}
    \caption{(A): Heat map of entanglement entropy at $N=6$. The Thirring model has three distinct phases, which can be clearly read off the diagram at $N=6$. (B): Heat map of entanglement entropy difference $\Delta S_{AB}$. (C): Heat map of teleported energy $\langle H_{n_B}\rangle$ at $N=6$. It is crucial that the value of the teleported energy peaks at the phase transition points, showing a clear correspondence to the phase diagram. (D) and (E): Time-evolution of entanglement entropy difference. This is due to the natural time evolution of the system, as seen when Bob does not perform any operations on his system after Alice's local operations. Decreasing $1-\frac{\Delta S_{AB}}{S_{AB}}$ in the early stages of time evolution means that entanglements broken by Alice's observations are recreated by the interactions in the system. (F): Simulation results of expected energy of Bob's local system obtained by QET. Error bars indicate statistical errors.}
    \label{fig:HV}
\end{figure*}
To facilitate clarity of results, we add a constant $\epsilon_i$ to every local Hamiltonian of the Thirring model
\begin{align}
\begin{aligned}
\label{eq:loc}
    H_\text{Th}&=\sum_{n}H_n
\end{aligned}
\end{align}
where $H_n$ is the local Hamiltonian including the nearest neighbor interactions and each $\epsilon_n$ should be chosen in such a way that 
\begin{align}
\begin{aligned}
    \bra{g}H_\text{Th}\ket{g}=\bra{g}H_n\ket{g}=0,~\forall i\in E
\end{aligned}    
\end{align}
where $\ket{g}$ is the ground state of the total Hamiltonian $H_\text{Th}$. Note that, in general, $\ket{g}$ is not the ground state of local $H_n$. The explicit form of Bob's local Hamiltonian and the details of computation are given in Supplemental Information (eq.~\eqref{eq:Bob_Hamiltonian}). It is important that non-trivial local manipulations, including measurement of the ground state, yield excited states and thus increase the energy expectation value. The increase in energy is supplied by the experimental apparatus. Moreover, our ground state $\ket{g}$ is an entangled state in general. 

The QET protocol is as follows. First, Alice measures her Pauli operator $\sigma_{n_A}$ by $P_{n_A}(\mu)=\frac{1}{2}(1+\mu \sigma_{n_A})$ and obtains either $\mu=-1$ or $+1$. Local measurement of the quantum state at a subsystem $A$ destroys this ground state entanglement. At the same time, energy $E_A$ from the device making the measurement is injected into the entire system. The injected energy $E_A$ is localized around the subsystem $A$ in the very early stages of time-evolution, however, it is not possible for Alice to extract $E_A$ from the system by her  operations alone at $n_A$. This is because information about $E_A$ is also stored in remote locations other than $n_A$ due to the entanglement that exists prior to the measurement. In other words, Alice's energy $E_A$ can be partially extracted at any location other than $n_A$. Now let us consider taking advantage of the quantum many-body nature of the quantum many-body system to extract energy from a different location other than $n_A$. This can be accomplished by LOCC, as shown below.

Via a classical channel, Alice sends her measurement result $\mu$ to Bob, who applies an operation $U_{n_B}(\mu)$ to his qubit and measures his local operators $X_{n_B},Y_{n_B},Z_{n_B}$ independently. The density matrix $\rho_\text{QET}$ after Bob operates $U_{n_B}(\mu)$ to $P_{n_A}(\mu)\ket{g}$ is 
where $\rho_\text{QET}$ is 
\begin{equation}
\label{eq:rho_QET}
    \rho_\text{QET}=\sum_{\mu\in\{-1,1\}}U_{n_B}(\mu)P_{n_A}(\mu)\ket{g}\bra{g}P_{n_A}(\mu)U^\dagger_{n_B}(\mu). 
\end{equation}
Using $\rho_\text{QET}$, the expected local energy at Bob's local system is evaluated as $\langle E_{n_B}\rangle=\Tr[\rho_\text{QET}H_{n_B}]$, which is negative in general. Due to the conservation of energy, $E_B=-\langle E_{n_B}\rangle (>0)$ is extracted from the system by the device that operates $U_{n_B}(\mu)$~\cite{PhysRevD.78.045006}. In this way, Alice and Bob can transfer the energy of the quantum system only by operations on their own local system and classical communication (LOCC). Those are summarized in Fig.~\ref{fig:protocol}. 

It should be noted that the Thirring model is a relativistic field theory in performing QET, which could be a problem if the particle is massless since the speed of classical communication does not exceed the speed of light. We will consider a massive particle and assume that Bob can receive energy faster than the time evolution rate of the system. 

In what follows we give the details about the operations of Alice and Bob. We define $U_{n_B}(\mu)$ by 
\begin{equation}
    U_{n_B}(\mu)=\cos\theta I-i\mu\sin\theta\sigma_{n_B},
\end{equation}
where $\theta$ obeys 
\begin{align}
    \cos(2\theta)&=\frac{\xi}{\sqrt{\xi^2+\eta^2}}\\
    \sin(2\theta)&=-\frac{\eta}{\sqrt{\xi^2+\eta^2}}
\end{align}
where
\begin{align}
    \xi&=\bra{g}\sigma_{n_B}H\sigma_{n_B}\ket{g}\\
    \eta&=\bra{g}\sigma_{n_A}\dot{\sigma}_{n_B}\ket{g}
\end{align}
with $\dot{\sigma_{n_B}}=i[H,\sigma_{n_B}]$. The local Hamiltonian should be chosen so that $[H,\sigma_{n_B}]=[H_{n_B},\sigma_{n_B}]$. 
The average quantum state $\rho_\text{QET}$ is obtained after Bob operates $U_{n_{B}}(\mu)$ to $P_{n_A}(\mu)\ket{g}$. Then the average energy Bob measures is 
\begin{equation}
\label{eq:QET}
    \langle E_{n_B}\rangle=\Tr[\rho_\text{QET}H_{n_B}]=\frac{1}{2}\left[\xi-\sqrt{\xi^2+\eta^2}\right], 
\end{equation}
which is negative if $\eta\neq 0$. If there is no energy dissipation, the positive energy of $-\langle E_{n_B\rangle}$ is transferred to Bob's device after the measurement due to energy conservation. Based on the quantum circuit in Fig.~\ref{fig:protocol}, we performed a quantum simulation of QET for $N=6,10,14$ at $\Delta(g)=-0.2,a=0.2$ and results are shown in Fig.~\ref{fig:HV}~(F). Dashed lines correspond to exact results. In this work, we put Alice and Bob near the boundary $n_{A}=1, n_{B}=N-2$. Bob's local energy can be calculated by the explicit form of his local Hamiltonian given in eq.\eqref{eq:Bob_Hamiltonian}. The simulation results are given in Table~\ref{tab:processor} in Sec.~\ref{sec:gate} of Suplimental Information. 

We next study the entanglement entropy between two subsystems $A,B$ such that $A\cap B=\emptyset, A\cup B=\{1,2,\cdots,N\}$. Let $\rho$ be a density operator on the entire system $A\cup B$. Then the entanglement entropy between $A$ and $B$ are defined by
\begin{equation}
    S(\rho)=-\Tr_{A}(\rho_A\log\rho_A),
\end{equation}
where $\rho_A$ is defined by tracing out the Hilbert space of $B$: $\rho_A=\Tr_{B}\rho$. In this study we choose $\rho$ as the ground state $\ket{g}$ of the Hamiltonian $(\rho=\ket{g}\bra{g})$. Fig.~\ref{fig:HV}~(A) shows the entanglement entropy between the left and right half subsystems, i.e., $A=\{0,\cdots,\frac{N}{2}\}$. The figure exhibits sharp peaks at the critical points of phase transitions that can be understood by the phase diagram of chiral condensate in Fig.~\ref{fig:chiral} in Appendix~\ref{sec:ASP}. Fig.~\ref{fig:HV}~(C) shows the teleported energy $\Tr[\rho_\text{QET}H_{n_B}]$ to Bob's local system. It is significant that the teleported energy is enhanced along the critical points of the phase transition. This will be understood by a relation between Bob's energy $\Tr[\rho_\text{QET}H_{n_B}]$ and the entanglement entropy difference $\Delta S_{SA}$, which is shown in Fig.~\ref{fig:HV}~(B).

The change in entropy before and after the measurement by Alice can be evaluated as follows
\begin{align}
    \Delta S_{AB}=S_{AB}-\sum_{\mu}p_\mu S_{AB}(\mu)
\end{align}
where $p_\mu$ is the probability distribution of $\mu$,  $S_{AB}(\mu)$ is the entanglement entropy after the measurement, $\xi=\arctan\left(\frac{k}{h}\right)$. After Alice's post-measurement, the state is mapped to 
\begin{equation}
    \ket{A(\mu)}=\frac{1}{\sqrt{p_\mu}}P_{n_A}(\mu)\ket{g}. 
\end{equation}
Then $S_{AB}(\mu)$ is calculated with the density matrix $\ket{A(\mu)}\bra{A(\mu)}$. 

As discussed in ~\cite{2011arXiv1101.3954H,2009JPSJ...78c4001H}, $\Delta S_{AB}$ is bounded below by a function $f(\xi,\eta)$ in such a way that
\begin{equation}
    \Delta S_{AB}\ge f(\xi,\eta)E_B.
\end{equation}
This indicates that the transferring energy involves a commensurate consumption of entropy. Similar to the Maxwell Demon argument~\cite{PhysRevLett.100.080403,PhysRevA.56.3374}, Bob's conditional operations reduce the entropy of the local system. If Bob does nothing after Alice's measurement, Figs~\ref{fig:HV}~(D) and (E) illustrate how the entanglement entropy is recreated by the natural time evolution of the system. 
Moreover, the maximal energy that Bob would receive is bounded below by the difference in entropy:
\begin{equation}
    \max_{U_1(\mu)}E_B\ge h(\xi,\eta)\Delta S_{AB},
\end{equation}
where $h(\xi,\eta)$ is a certain function. 

Although it is difficult to analytically obtain the concrete forms of functions $f$ and $g$, the results of this study show that there is a clear correspondence between the energy obtained by QET and the phase diagram of QFT.

%%%%%%%%%%%%%%%%%%%%%%%%%%%%%%%%%%%%%%%%%%%%%%%%%%%%%%
\section*{Acknowledgement}
I thank Adrien Florio, David Frenklakh, Sebastian Grieninger, Fangcheng He, Masahiro Hotta, Dmitri Kharzeev, Yuta Kikuchi, Vladimir Korepin, Qiang Li, Adam Lowe, Ren\'{e} Meyer, Shuzhe Shi, Hiroki Sukeno, Tzu-Chieh Wei, Kwangmin Yu and Ismail Zahed for fruitful communication and collaboration. I thank Megumi Ikeda for providing the cartoons. I acknowledge the use of IBM quantum computers and simulators. I was supported by the U.S. Department of Energy, Office of Science, National Quantum Information Science Research Centers, Co-design Center for Quantum Advantage (C2QA) under Contract No.DESC0012704.
%%%%%%%%%%%%%%%%%%%%%%%%%%%%%%%%%%%%%%%%%%%%%%%%%%%%%%%%%%%%%%%%%

\section*{Author contribution}
All work was performed by the author.

\section*{Competing interests}
The author declares that there is no competing financial interests.

\appendix
%%%%%%%%%%%%%%%%%%%%%%%%%%%%%%%%%%%%%%
\section{\label{sec:gate}Quantum Gates and Measurement}

\begin{table*}
    \centering
    \begin{tabular}{|c|c|c|c|c|c|c|c|}\toprule
      \multicolumn{2}{|c|}{$m$} &0.5 &1 & 1.5 & 2  \\\hline
       \multirow{3}{*}{$\langle Z_{N-2}\rangle$}&$N=14$& $0.2303\pm0.0010$ & $0.3459\pm0.0009$ & $0.4111\pm0.0009$ &$0.5050\pm0.0009$ \\&$N=10$& $0.2453\pm0.0010$ & $0.3021\pm0.0010$ & $0.4125\pm0.0009$ &$0.5069\pm0.0009$ \\
       &$N=6$& $0.2132\pm0.0010$ & $0.3245\pm0.0010$ & $0.4256\pm0.0009$ &$0.5141\pm0.0009$ \\
       %&\text{ibm\_cairo}& $0.2132\pm0.0010$ & $0.3245\pm0.0010$ & $0.4256\pm0.0009$ &$0.5141\pm0.0009$ \\
       \hline
       \multirow{3}{*}{$\langle X_{N-3}X_{N-2}\rangle$}&$N=14$& $0.5281\pm0.0008$ & $0.5211\pm0.0009$ & $0.5710\pm0.0008$ &$0.5550\pm0.0008$ \\&$N=10$& $0.5376\pm0.0008$ & $0.5717\pm0.0008$ & $0.5697\pm0.0008$&$0.5535\pm0.0008$ \\
       &$N=6$& $0.5270\pm0.0008$& $ 0.5454\pm0.0008$ & $0.5522\pm0.0008$ &$0.5436\pm0.0008$ %\\&\text{ibm\_cairo}& $0.2132\pm0.0010$ & $0.3245\pm0.0010$ & $0.4256\pm0.0009$ &$0.5141\pm0.0009$ 
       \\\hline
       \multirow{3}{*}{$\langle X_{N-2}X_{N-1}\rangle$}&$N=14$& $0.7977\pm0.0006$ & $0.7572\pm0.0007$ & $0.6757\pm0.0007$ &$0.6304\pm0.0008$ \\&$N=10$& $0.7813\pm0.0006$ & $0.7271\pm0.0007$ & $0.6776\pm0.0007$&$0.6308\pm0.0008$ \\
       &$N=6$& $0.8005\pm0.0006$& $0.7412\pm0.0007$ & $0.6840\pm0.0007$ &$0.6330\pm0.0008$ \\\hline
       \multirow{3}{*}{$\langle Y_{N-3}Y_{N-2}\rangle$}&$N=14$& $0.5287\pm0.0008$ & $0.5218\pm0.0009$ & $0.5712\pm0.0008$ &$0.5548\pm0.0008$ \\&$N=10$& $0.5375\pm0.0008$ & $0.5728\pm0.0008$ & $0.5687\pm0.0008$ &$0.5541\pm0.0008$ \\
       &$N=6$&$0.5334\pm0.0008$& $0.5531\pm0.0008$ & $0.5579\pm0.0008$ &$0.5502\pm0.0008$ \\
       %&\text{ibm\_cairo}& $0.2132\pm0.0010$ & $0.3245\pm0.0010$ & $0.4256\pm0.0009$ &$0.5141\pm0.0009$ \\
       \hline
       \multirow{3}{*}{$\langle Y_{N-2}Y_{N-1}\rangle$}&$N=14$& $0.7958\pm0.0006$ & $0.7578\pm0.0006$ & $0.6763\pm0.0007$ &$0.6302\pm0.0008$ \\&$N=10$& $0.7811\pm0.0006$ & $0.7270\pm0.0007$ & $0.6773\pm0.0008$&$0.6301\pm0.0008$ \\
       &$N=6$& $0.8018\pm0.0006$ & $0.7427\pm0.0007$ & $0.6849\pm0.0007$ &$0.6319\pm0.0008$
       %\\&\text{ibm\_cairo}& $0.2132\pm0.0010$ & $0.3245\pm0.0010$ & $0.4256\pm0.0009$ &$0.5141\pm0.0009$ 
       \\\hline
       \multirow{3}{*}{$\langle Z_{N-3}Z_{N-2}\rangle$}&$N=14$& $-0.2173\pm0.0010$ & $-0.2426\pm0.0010$ & $-0.4943\pm0.0009$ &$-0.5658\pm0.0008$ \\&$N=10$& $-0.2026\pm0.0010$ & $-0.4092\pm0.0009$ & $-0.4957\pm0.0009$&$-0.5663\pm0.0008$ \\
       &$N=6$& $-0.2764\pm0.0010$& $-0.3749\pm0.0009$ & $-0.4724\pm0.0009$ &$-0.5569\pm0.0008$ %\\&\text{ibm\_cairo}& $0.2132\pm0.0010$ & $0.3245\pm0.0010$ & $0.4256\pm0.0009$ &$0.5141\pm0.0009$ 
       \\\hline
       \multirow{3}{*}{$\langle Z_{N-2}Z_{N-1}\rangle$}&$N=14$& $-0.6429\pm0.0008$ & $-0.6646\pm0.0007$ & $-0.7268\pm0.0007$ &$-0.7637\pm0.0006$ \\&$N=10$& $-0.5915\pm0.0008$ & $-0.6934\pm0.0007$ & $-0.7283\pm0.0006$ &$-0.7661\pm0.0006$ \\
       &$N=6$& $-0.7083\pm0.0007$& $-0.7196\pm0.0007$ & $-0.7413\pm0.0007$ &$-0.7723\pm0.0006$ %\\&\text{ibm\_cairo}& $0.2132\pm0.0010$ & $0.3245\pm0.0010$ & $0.4256\pm0.0009$ &$0.5141\pm0.0009$ 
       \\\hline
    \end{tabular}
    \caption{Expectation values of operators evaluated by $10^6$ sampling data with a simulator. $\Delta(g)$ and the lattice spacing $a$ are fixed to $-0.2$ and $0.2$. By substituting those values into Bob's expected energy $\langle H_{n_B}\rangle$, one can recover the result of Fig.~\ref{fig:HV} (F).}
    \label{tab:processor}
\end{table*}
\begin{figure*}
    \centering
    \includegraphics[width=\linewidth]{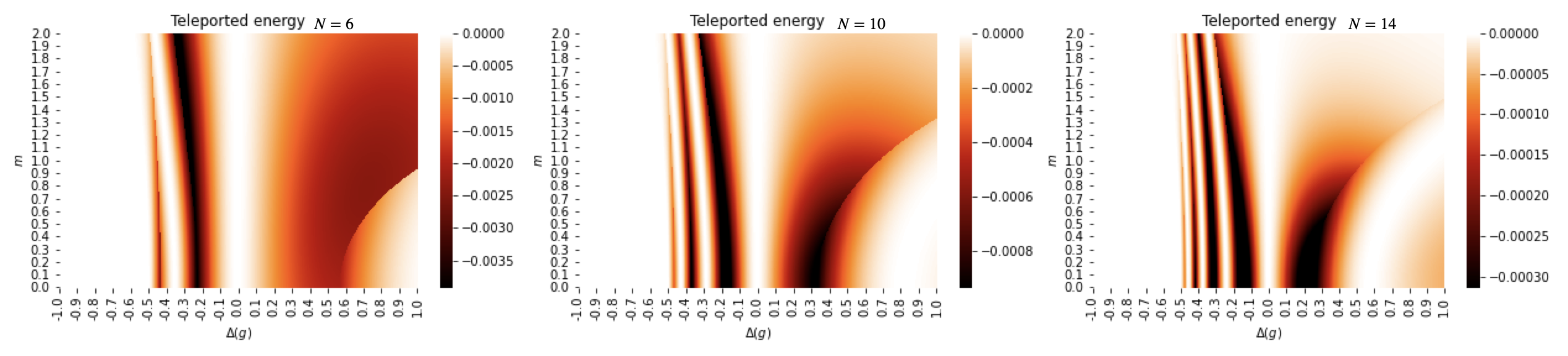}
    \caption{Teleported energy for a different system size $N=6,10,14$.}
    \label{fig:my_label}
\end{figure*}

The goal of this section is to describe how to compute Bob's local energy $\langle H_{n_B}\rangle$ gained by quantum energy teleportation, using the quantum circuit shown in Fig.~\ref{fig:protocol}. For this we provide a self-contained description of the background knowledge used in the main text. We use the following one-qubit operators whose matrix representations are given as 
\begin{equation}
\begin{aligned}
    X&=\begin{pmatrix}
        0&1\\
        1&0
    \end{pmatrix},
    Y=\begin{pmatrix}
        0&-i\\
        i&0
    \end{pmatrix},
    Z=\begin{pmatrix}
        1&0\\
        0&-1
    \end{pmatrix},\\
    S&=\begin{pmatrix}
        1&0\\
        0&i
    \end{pmatrix},
    H=\frac{1}{\sqrt{2}}\begin{pmatrix}
        1&1\\
        1&-1
    \end{pmatrix}.
\end{aligned}
\end{equation}
We use $\ket{0}=\binom{1}{0},\ket{1}=\binom{0}{1}$ for the computational basis states, which are eigenstates of $Z$: $Z\ket{0}=\ket{0},Z\ket{1}=-\ket{1}$. We also work with another basis vectors $\ket{\pm}=\frac{\ket{0}\pm\ket{1}}{\sqrt{2}}$. They are eignestates of $X$: $X\ket{-}=-\ket{-},X\ket{+}=-\ket{+}$. Note that $\ket{\pm}$ are created by applying $H$ to $\ket{0}$ and $\ket{1}$; $H\ket{0}=\ket{+},H\ket{1}=\ket{-}$. For example, Alice finds $\mu=\pm1$ by observing the eigenvalues $\pm1$ of her local Pauli $X$ operator. 

The rotation of $X,Y,Z$ is defined by 
\begin{equation}
    R_X(\alpha)=e^{-i\frac{\alpha}{2} X},~R_Y(\alpha)=e^{-i\frac{\alpha}{2} Y},~R_Z(\alpha)=e^{-i\frac{\alpha}{2} Z}.
\end{equation}

We use two-qubit gate operations. In general, a control $U$ operation $\Lambda(U)$ is defined by 
\begin{equation}
    \Lambda(U)=\ket{0}\bra{0}\otimes I+\ket{1}\bra{1}\otimes U
\end{equation}
and the corresponding diagram is drwan as 
\begin{figure}[H]
    \centering
\begin{quantikz}
\midstick[2,brackets=none]{control $U$=}\qw&\ctrl{1}&\qw\\
\qw&\gate{U}&\qw
\end{quantikz}
\end{figure}

One of the most frequently used controlled gates is a CNOT gate $\text{CNOT}=\Lambda(X)$, whose diagram is especially drawn as
\begin{figure}[H]
    \centering
\begin{quantikz}
\midstick[2,brackets=none]{CNOT=}\qw&\ctrl{1}&\qw\\
\qw&\targ{}&\qw
\end{quantikz}
\end{figure}

It is convenient to define an anti-control gate, which is activated when the control bit is in state $\ket{0}$: $\ket{1}\bra{1}\otimes I+\ket{0}\bra{0}\otimes U$, whose diagram is drawn as  
\begin{figure}[H]
    \centering
\begin{quantikz}
\midstick[2,brackets=none]{Anti-control $U$=}\qw&\octrl{1}&\qw\\
\qw&\gate{U}&\qw
\end{quantikz}
=
\begin{quantikz}
\gate{X}&\ctrl{1}&\gate{X}\\
\qw&\gate{U}&\qw
\end{quantikz}
\end{figure}

With those operators, we can draw time evolution of $XX,YY,ZZ$ type interactions of spins as  
\begin{widetext}
\begin{figure}[H]
    \centering
\begin{quantikz}
\midstick[2,brackets=none]{$e^{-i\frac{\alpha}{2} (X_nX_{n+1}+Y_nY_{n+1})}$=}\qw&\ctrl{1}&\gate{H}&\ctrl{1} &\gate{R^{(n)}_Z(\alpha)}&\ctrl{1}&\gate{H}&\ctrl{1}&\qw\\
\qw&\targ{}&\qw&\targ{}&\gate{R^{(n+1)}_Z(-\alpha)}&\targ{}&\qw&\targ{}&\qw
\end{quantikz}
\begin{quantikz}
\midstick[2,brackets=none]{$e^{-i\frac{\alpha}{2} Z_nZ_{n+1}}$=}\qw&\ctrl{1} &\qw&\ctrl{1}&\qw\\
\qw&\targ{}&\gate{R^{(n+1)}_Z(\alpha)}&\targ{}&\qw
\end{quantikz}
\end{figure}
\end{widetext}
A Hamiltonian $H_Z=\sum_{n=1}^Na_nZ_n$ containing only local $Z_n$s is implemented by 
\begin{equation}
    e^{-i\alpha H_Z}=\prod_{n=1}^NR^{(n)}_Z(2a_n\alpha),
\end{equation}

Now we describe the measurement of quantum operators. Measurement of $Z_n$ is done by the following circuit
\begin{figure}[H]
    \centering
    \begin{quantikz}
&\meter{}
\end{quantikz}
\end{figure}
The output of the measurement is a bit string $b_n\in\{0,1\}$. Since the eigenvalues of $Z$ are $-1,1$, we convert the bit string into $1-2b_n$. Let $n_\text{shot}$ be the number of repetitions of the circuit, and $\text{counts}_{b_0\cdots b_{N-1}}$ be the number of times $b_0,\cdots, b_{N-1}$ are detected. Therefore $\frac{\text{counts}_{b_0\cdots b_{N-1}}}{n_\text{shots}}$ is the probability that a bit string of $b_0\cdots b_{N-1}$ is obtained. Then the expectation value of $Z_n$ is computed by the formula
\begin{align}
 \langle Z_n\rangle&=\sum_{b_0,\cdots,b_{N-1}}(1-2b_n)\frac{\text{counts}_{b_0\cdots b_{N-1}}}{n_\text{shots}}.
\end{align}

Measurement of $X_nX_{n+1}$ is done by the following circuit
\begin{figure}[H]
    \centering
    \begin{quantikz}
&\gate{H}&\meter{}\\
&\gate{H}&\meter{}
\end{quantikz}
\end{figure}
Note that $H$ maps $\ket{0},\ket{1}$ to $\ket{+},\ket{-}$, which are eigenvectors of $X$. The output is again a bit string $b_nb_{n+1}\in\{00,01,10,11\}$. They are converted to the eigenvalues of $X_nX_{n+1}$ by $(1-2b_n)(1-2b_{n+1})$. Then the expectation value of $X_nX_{n+1}$ is computed by the formula
\begin{align}
 \langle X_nX_{n+1}\rangle&=\sum_{b_n,b_{n+1}}(1-2b_n)(1-2b_{n+1})\frac{\text{counts}_{b_{n}b_{n+1}}}{n_\text{shots}}.
\end{align}

Similarly, measurement of $Y_{n}Y_{n+1}$ is possible by the following circuit
\begin{figure}[H]
    \centering
    \begin{quantikz}
&\gate{S^\dagger}&\gate{H}&\meter{}\\
&\gate{S^\dagger}&\gate{H}&\meter{}
\end{quantikz}
\end{figure}
in such a way that
\begin{align}
 \langle Y_nY_{n+1}\rangle&=\sum_{b_n,b_{n+1}}(1-2b_n)(1-2b_{n+1})\frac{\text{counts}_{b_nb_{n+1}}}{n_\text{shots}}.
\end{align}

Now let us discuss how to compute the expectation energy of the massive Thirring model~(eqs.~\eqref{eq:Ham_spin} and \eqref{eq:loc}). Once we obatin $\langle Z_n\rangle,\langle Z_nZ_{n+1}\rangle,\langle X_{n}X_{n+1}\rangle,\langle Y_nY_{n+1}\rangle$ as shown in Table~\ref{tab:processor}, then we can compute the expectation value of teleported energy by 
\begin{align}
\begin{aligned}
H_\text{Th}&=\sum_{n=0}^{N-2}H_n\\
\langle H_{n}\rangle&=\langle H_{\pm,n}\rangle+\langle H_{m,n}\rangle+\langle H_{ZZ,n}\rangle+\epsilon.
\end{aligned}
\end{align}
For example Bob's local Hamiltonian $H_{n_B}$ is given as follows
\begin{align}
\begin{aligned}
\label{eq:Bob_Hamiltonian}
H_{n_B}=&H_{\pm,B}+H_{m,B}+H_{ZZ,B}+\epsilon_n\\
    H_{\pm,B}&=-\frac{1}{4a}[X_{N-2}(X_{N-3}+X_{N-1})\\
    &+Y_{N-2}(Y_{N-3}+Y_{N-1})],\\
    H_{m,B}=&\frac{m}{2}(-1)^{N-1}Z_{N-2}\\
    H_{ZZ,B}=&\frac{\Delta(g)}{8}Z_{N-2}(Z_{N-3}+Z_{N-1})+\frac{\Delta(g)}{2a}Z_{N-2}.
\end{aligned}
\end{align}
Now it will be clear that $[H_\text{Th},\sigma_{n_B}]=[H_{n_B},\sigma_{n_B}]$ since $\sigma_{n_B}$ is a local operator. The teleported energy $\langle H_{n_B}\rangle$ to Bob's local system is shown in Fig.~\ref{fig:HV} (F) in the main text. Bob will receive $-\langle H_{n_B}\rangle$ through his measurement device.

\section{Sine-Gordon Model and Thirring Model}\label{sec:quench}
The sine-Gordon model is a theory of a single scalar field $\phi(x)$ whose Lagrangian is written as
\begin{equation}
\label{eq:Lag}
    \mathcal{L}_\text{SG}=\frac{1}{2}\partial_\mu\phi(x)\partial^\mu\phi(x)+\frac{\alpha}{\beta^2}(\cos(\beta\phi(x))-1),
\end{equation}
where $\alpha$ and $\beta$ are real positive parameters.
This model is invariant under 
\begin{equation}
    \phi(x)\to\phi'(x)=\phi(x)+\frac{2\pi n}{\beta},~n\in\mathbb{Z}. 
\end{equation}

The Thirring model is a theory of a self–coupled Dirac field
\begin{equation}
    \mathcal{L}_\text{Th}=\overline{\psi}(i\gamma^\mu\partial_\mu-m)\psi-\frac{g}{2}\overline{\psi}\gamma^\mu\psi\overline{\psi}\gamma_\mu\psi,
\end{equation}
where $m$ is the fermion mass, $g$ is  the dimensionless four-fermion coupling constant and $\psi=\psi(x)$ is a spinor field with two components $\psi_1(x)$ and $\psi_2(x)$. It is widely known that the massive Thirring model is dual to the sine-Gordon model and the classical two-dimensional $XY$ model~\cite{PhysRevD.11.3026,2001EPJC...20..723F}. For example a Kosterlitz-Thouless phase transition at $T\sim K\pi/2$ in the $XY$ model corresponds to a critical point $g\sim -\pi/2$, called Coleman’s instability point, in the Thirring model. They are also related with a critical point at $t\sim 8\pi$ in the sine-Gordon model. 

The model is obviously invariant under $U(1)_V$ group 
\begin{equation}
    \psi(x)\to \psi'(x)=e^{i\alpha_V}\psi(x).
\end{equation}
The corresponding conserved current is 
\begin{equation}
    j_\mu=\overline{\psi}\gamma_\mu\psi. 
\end{equation}

If $m=0$, the Thirring model is also invariant under the chiral group $U(1)_V\times U(1)_A$
\begin{align}
    \begin{aligned}
    \psi(x)&\to \psi'(x)=e^{i\alpha_V}\psi(x)\\
    \psi(x)&\to \psi'(x)=e^{i\alpha_A\gamma_5}\psi(x),
    \end{aligned}
\end{align}
under which the pseudo-vector current
\begin{equation}
    j^5_\mu=\epsilon_{\mu\nu}j_\nu~~(\epsilon_{01}=1-\epsilon_{10}=1)
\end{equation}
also conserves $\partial_\mu j^5_{\mu}=0$. 

Bosonization of the interaction term is described by 
\begin{align}
    \begin{aligned}
    \mathcal{L}^\text{SG}_\text{int}(x)&=\frac{\alpha}{\beta^2}\cos(\beta\phi(x))=\frac{\alpha}{2\beta^2}(A_+(x)+A_-(x))\\
    \mathcal{L}^\text{Th}_\text{int}(x)&=-m\overline{\psi}(x)\psi(x)=-m(\sigma_+(x)+\sigma_-(x)),
    \end{aligned}
\end{align}
where $A_\pm(x)=e^{\pm i \beta\phi(x)}$ and $\sigma_\pm(x)=\frac{1}{2}\overline{\psi}(x)(1\pm\gamma^5)\psi(x)$

According to the Coleman's prescription, those two models are related with
\begin{align}
\begin{aligned}
    \overline{\psi}(x)\psi(x)&=\frac{-M\cos(\beta\phi(x))+m}{g}\\
    i\overline{\psi}(x)\gamma^5\psi(x)&=-\frac{M\sin(\beta\phi(x))}{g}\\
    m\overline{\psi}(x)\left(\frac{1\mp\gamma^5}{2}\right)\psi(x)&=-\frac{\alpha}{2\beta^2}e^{\pm i\beta\phi(x)}+\frac{m^2}{2g}. 
\end{aligned}
\end{align}
The massive Thirring model and the sine-Gordon model are equivalent only if $\beta^2<8\pi$. 

\section{\label{sec:Ham}Hamiltonian Formalism of Thirring Model}
The Hamiltonian including quantum effects in the energy-momentum tensor at the operator level can be written as 
\begin{align}
\begin{aligned}
    H_{\text{Th}}=\int dx\bigg[&-iZ_{\psi}(g)\bar{\psi}\gamma^1\partial_1\psi+m\bar{\psi}\psi\\
    &+\frac{g}{4}(\bar{\psi}\gamma^0\psi)^2-\frac{\tilde{g}}{2}(\bar{\psi}\gamma^1\psi)^2\bigg],
\end{aligned}
\end{align}
where $Z_\psi(g)$ is the wavefunction renormalization constant and $\tilde{g}=\frac{g}{2}\left(1+\frac{2g}{\pi}\right)^{-1}$~\cite{hagen1967new,PhysRevD.4.1635,PhysRevD.7.550}. 
\begin{equation}
    \psi_1(x)\to\frac{1}{\sqrt{a}}\chi_{2n},~\psi_2(x)\to\frac{1}{\sqrt{a}}\chi_{2n+1}
\end{equation}
We address this Hamiltonian on a lattice one-dimensional lattice with open boundary condition. Then the discretized Hamiltonian is 
\begin{align}
\begin{aligned}
    H_\text{Th}=&-\frac{i}{2a}Z_{\psi}(g)\sum_{j=1}^{N-1}\left(\chi^\dagger_j\chi_{j+1}+h.c.\right)\\
    &+m\sum_{j=1}^{N}(-1)^nn_j+\frac{{g}}{2a}\sum_{j=1}^{N-1}n_jn_{j+1},
\end{aligned}
\end{align}
where $n_j=\chi^\dagger_j\chi_j$ is the fermion number operator. 

For the purpose of quantum computation, we convert the Hamiltonian into the corresponding spin Hamiltonian. We will work with 
\begin{equation}
    \gamma^0=X,~\gamma^1=-iY,~\gamma^5=-iXY
\end{equation}

We first write down the discrete Hamiltonian by means of staggered fermion. Based on the established method~\cite{PhysRevB.14.2153,alcaraz1995critical}, we can replace $Z_\psi(g)$ and ${g}$ with
\begin{align}
\begin{aligned}
    Z_\psi(g)&\to\frac{\pi-g}{\pi\sin\left(\frac{\pi-g}{2}\right)}\\
    {g}&\to\frac{2(\pi-g)}{\pi}\cot\left(\frac{\pi-g}{2}\right)
\end{aligned}
\end{align}
As a result we obtain the Hamiltonian of the massive Thirring model we used in this work
\begin{equation}
    H_\text{Th}=\frac{2\gamma}{a\pi\sin(\gamma)}H,
\end{equation}
where the explicit representation of $H$ is 
\begin{align}
\begin{aligned}
H&=-\frac{i}{2a}\sum_{n=1}^{N-1}(\chi^\dagger_n\chi_{n+1}-\chi^\dagger_{n+1}\chi_{n})\\
&+m\sum_{n=1}^{N}(-1)^n \chi^\dagger_n\chi_n+\frac{\Delta(g)}{a}\sum_{n=1}^{N-1}\chi^\dagger_n\chi_n\chi^\dagger_{n+1}\chi_{n+1},
\end{aligned}
\end{align}
where $\Delta(g)=\cos\left(\frac{\pi-g}{2}\right)$
\cite{PhysRevD.11.3026,PhysRevB.14.2153,PhysRevD.11.2088,2018slft.confE.229B,2019PhRvD.100i4504B,2017arXiv171009993B}

The Jordan-Wigner transformation maps fermionic operators to spin operators~\cite{Jordan:1928wi}. It is commonly used in the study of quantum many-body systems and plays a key role in the study of strongly correlated electron systems. After the Jordan-Winger transformation
\begin{align}
\begin{aligned}
 \chi_n&=\frac{X_n-iY_n}{2}\prod_{m=1}^{n-1}(-i Z_m),\\
 \chi^\dag_n&=\frac{X_n+iY_n}{2}\prod_{i=m}^{n-1}(i Z_m),
\end{aligned}
\end{align}
we arrive at the spin representation of the massive Thirring model
\begin{align}
\begin{aligned}
H_\text{spin}&=-\frac{1}{4a}\sum_{n=1}^{N-1}(X_nX_{n+1}+Y_nY_{n+1})\\
&+\frac{m}{2}\sum_{n=1}^{N}(-1)^n(Z_n+1)\\
&+\frac{\Delta(g)}{a}\sum_{n=1}^{N-1}\left(\frac{Z_n+1}{2}\right)\left(\frac{Z_{n+1}+1}{2}\right).
\end{aligned}
\end{align}

Those correspondences are summarized in the following dictionary.
\begin{table}[H]
\begin{center}
\begin{tabular}{c|c|c}\toprule
Dirac & Staggerd  & Pauli \\\hline
     $\overline{\psi}\psi$ & $\frac{(-1)^n}{a}\chi^\dagger_n\chi_n$ &  $\frac{(-1)^n}{2a}(Z_n+1)$ \\
     $\overline{\psi}\gamma_0\psi$ & $\frac{1}{a}\chi^\dagger_n\chi_n$ &  $\frac{1}{2a}(Z_n+1)$ \\
     $\overline{\psi}\gamma_1\psi$ & $\frac{1}{2a}(\chi^\dagger_n\chi_{n+1}+\chi^\dagger_{n+1}\chi_{n})$ &  $\frac{1}{4a}(X_nY_{n+1}-Y_nX_{n+1})$ \\
    $\overline{\psi}\gamma_5\psi$ & $\frac{(-1)^n}{2a}(\chi^\dagger_n\chi_{n+1}-\chi^\dagger_{n+1}\chi_{n})$ &  $-\frac{i(-1)^n}{4a}(X_nX_{n+1}+Y_nY_{n+1})$ \\
    $\overline{\psi}\gamma_1\partial_1\psi$ & $-\frac{1}{2a^2}(\chi^\dagger_n\chi_{n+1}-\chi^\dagger_{n+1}\chi_{n})$ &  $-\frac{i}{4a^2}(X_nX_{n+1}+Y_nY_{n+1})$ \\
\end{tabular}
\end{center}
    \caption{Correspondence among the three different representations of fermionic fields.}
    \label{tab:dic}
\end{table}

\section{\label{sec:ASP}Adiabatic Ground State Preparation}
\label{sec:quantum_simulation}
\subsection{General Remark}
One of the most difficult problems in performing quantum computation is how to design an initial state since it can affect the overall outcome of the computation. A good initial state is one that is easy to prepare and that allows for efficient quantum gates and measurements to be applied to it. One common initial state used in quantum computation is the ground state of a Hamiltonian. However a difficulty in preparing a good initial state is the quality of the qubits and the limitation of gate depth. If the qubits are prone to errors, initializing them with high fidelity is very challenging. A commonly used method to solve such problems is the adiabatic quantum computation (AQC), which is a model of quantum computation based on the adiabatic theorem of quantum mechanics. It can be used to solve certain optimization problems by encoding the problem into the energy levels of a time-dependent Hamiltonian, and then slowly varying the Hamiltonian over time to drive the system through a series of states that correspond to the solution of the problem. One such quantum computation method is quantum annealing, which is a heuristic optimization method that is used to find the global minimum of a given spin Hamiltonian~\cite{PhysRevE.58.5355} and practically used for various purposes~\cite{2019NatSR...912837I,ohzeki2020breaking,kadowaki2023greedy}. 

A well-known example is the Quantum Adiabatic Algorithm (QAA), it is based on the Hamiltonian of the problem to be solved, the initial Hamiltonian is chosen such that its ground state is easy to prepare and known and the final Hamiltonian is such that its ground state is the solution of the problem. The system is then slowly evolved to the final Hamiltonian and the final state is the ground state of the final Hamiltonian. One of the main requirements for the QAA to work correctly is that the adiabatic condition holds, this condition states that the adiabatic evolution should be slow enough so that the system remains in the ground state at all times, if this condition is not met the system may end up in excited state and the final solution will not be correct. One challenge that arises during AQC is the trade-off between the speed of the evolution and the accuracy of the final solution. The adiabatic theorem states that the system remains in the ground state if the evolution is slow enough. However, if the evolution is too slow, the computation may become infeasible due to the long running time. On the other hand, if the evolution is too fast, the system may not stay in the ground state and the final solution may be inaccurate.

Another challenge is the presence of a phase transition, which is a point where the ground state of the Hamiltonian changes. At these points, the energy gap between the ground state and the first excited state can become small, which makes it more difficult to maintain the system in the ground state. This can also lead to critical slowing down, in which the system slows down drastically as it evolves through the phase transition point. In the context of adiabatic quantum computation, a phase diagram could be used to understand how the performance of the computation changes as a function of parameters such as the duration of the computation or the strength of the interactions between the quantum particles.

Additionally, AQC is also sensitive to noise and errors in the system, as small fluctuations can cause the system to leave the ground state. This can lead to incorrect solutions or a loss of quantum coherence.

\subsection{Method}
Here we describe a way to prepare the ground state by a quantum computer. For this, we use the method so-called  adiabatic state preparation. The essence of the method for studying phase transitions with adiabatic quantum calculations has already been established in many cases with quantum annealing~\cite{PhysRevE.58.5355,2020QuIP...19..331I,PhysRevE.104.024136}. 
We first prepare a state $\ket{\text{vac}_0}$ which is a known ground state of an initial Hamiltonian $H_\text{initial}$. We use the time-dependent Hamiltonian $H(t)$ which is equal to the target Hamiltonian $H_\text{target}$ at the end of computation: 
\begin{equation}
\label{eq:Ht}
    H(t)=tH_\text{target}+(1-t)H_\text{initial}.
\end{equation}
 By adiabatically changing parameter $t$, we expect to obtain the ground state of the target Hamiltonian 
\begin{equation}
\label{eq:integral}
\ket{\text{vac}}=\lim_{T\to\infty}\mathcal{T}\exp\left(-i\int_0^TdtH(t)\right)\ket{\text{vac}}_0.
\end{equation}

We have two reasonable choices of initial Hamiltonian. The first one is 
\begin{equation}
\label{eq:initHam_m}
    H_m=\frac{m}{2}\sum_{n=1}^N(-1)^n Z_n,~~m>0
\end{equation}
whose ground state is $\ket{\text{vac}}_0=\ket{0101\cdots01}=\prod_{i=1}^{N/2}X_{2i}\ket{0\cdots0}$, with $Z\ket{0}=\ket{0}$ and $Z\ket{1}=-\ket{1}$. The other choice is 
\begin{equation}
\label{eq:initHam_g}
    H_g=\frac{\Delta(g)}{a}\sum_{n=1}^{N-1}\left(\frac{Z_n+1}{2}\right)\left(\frac{Z_{n+1}+1}{2}\right),~~\Delta(g)<0
\end{equation}
whose ground state is $\ket{\text{vac}}_0=\ket{0\cdots0}$. Each of those states can be easily prepared but we will choose one of them so that any possible errors including become small. 

The matrix representation of the chiral condensate is given as 
\begin{equation}
    \sum_{n=1}^N\bar{\psi}\psi_n=\frac{1}{2a}\sum_{n=1}^N(-1)^nZ_n. 
\end{equation}
Its vacuum expectation value is $-\frac{N}{2a}$ with the ground state $\ket{0101\cdots01}$ of an initial Hamiltonian~\eqref{eq:initHam_m}
\begin{align}
    \bra{01\cdots01}\bar{\psi}\psi\ket{01\cdots01}=-\frac{N}{2a}
\end{align}
Those two states corresponds to the ground states of the Hamiltonian~\eqref{eq:Ham_spin} in the limit: $\lim_{m\to\infty}\frac{1}{m}(H_\text{spin}-H_m)=0$ and $\lim_{|\Delta(g)|\to\infty}\frac{1}{|\Delta(g)|}(H_\text{spin}-H_g)=0$, respectively. We chose an initial state and parameters are chosen so that the effect of the phase transition on the probability distribution and statistical errors are relatively small. In the rest of this section, we describe how we manage an adiabatic state preparation. 

\begin{figure}[H]
    \centering
    \includegraphics[width=\linewidth]{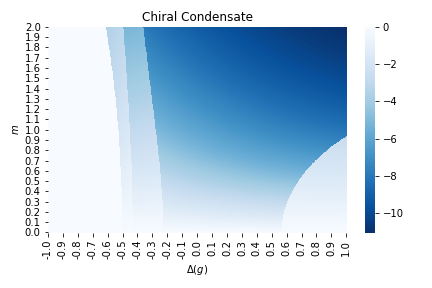}
    \caption{Chiral condensate}
    \label{fig:chiral}
\end{figure}

In the adiabatic process, we decompose the time-dependent Hamiltonian $H(t)=H_\text{static}+H_\text{dynamic}(t)$ into the static part and the dynamic part. For example, if the Hamiltonian \eqref{eq:initHam_m} is used for the initial Hamiltonian, the static part is 
\begin{equation}
\label{eq:Hz_static}
    H_\text{static}=\frac{m}{2}\sum_{n=1}^N(-1)^nZ_n
\end{equation}
and the dynamic part is 
\begin{align}
\begin{aligned}
\label{eq:Hzz_dynamic}
    H_\text{dynamic}(t)=&-\frac{w(t)}{4a}\sum_{n=1}^{N-1}(X_nX_{n+1}+Y_nY_{n+1})\\
    &+\frac{g(t)}{a}\sum_{n=1}^{N-1}\left(\frac{Z_n+1}{2}\right)\left(\frac{Z_{n+1}+1}{2}\right)
\end{aligned}
\end{align}
with
\begin{align}
\begin{aligned}
    \Delta(g)(t)&=\frac{t}{T}\left(\frac{\Delta(g_\text{target})t}{T}+\Delta(g_0)\left(1-\frac{t}{T}\right)\right),\\
    w(t)&=\frac{t}{T},
    \end{aligned}
\end{align}
where $\Delta(g_0)$ is an initial coupling that should be chosen in a way that statistical errors become small, and $\Delta(g_\text{target})$ is the target coupling parameter.

\subsection{Study on Phase Diagram by Adiabatic State Preparation}
\begin{figure}[H]
    \centering
    \includegraphics[width=\linewidth]{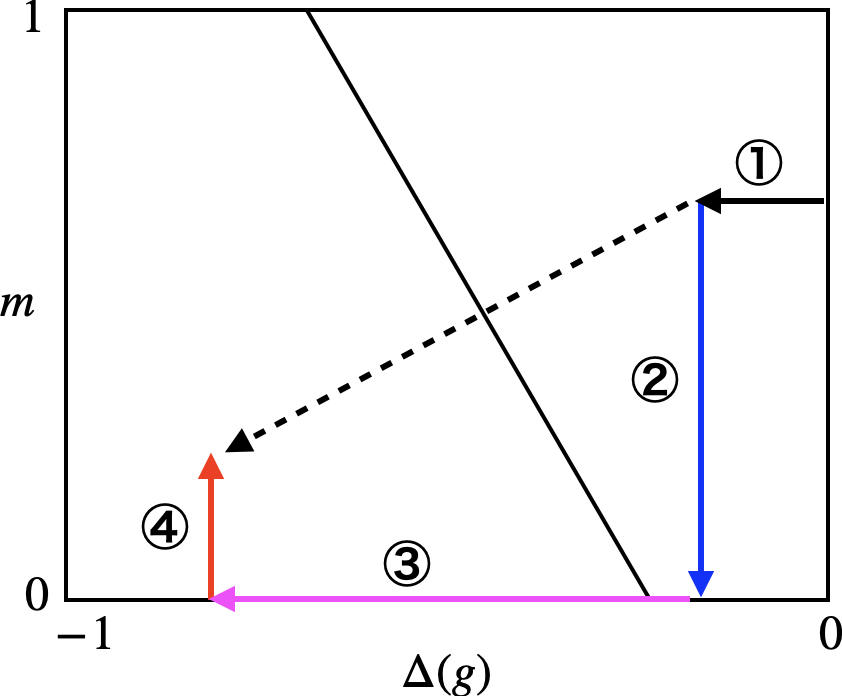}
    \caption{Schematic design of path to compute the vacuum expectation value of chiral condensate by adiabatic state preparation. Paths should be selected so that results are not affected by phase transitions.}
    \label{fig:path}
\end{figure}

It will be interesting to explore the phase diagram of the massive Thirring model by the adiabatic algorithm, in particular for the purpose of quantum computation since we have to prepare the ground state anyway. By using two different initial Hamiltonians above, one can draw the phase diagram of the massive Thirring model. In Fig.~\ref{fig:path}, we illustrate a path we take for the adiabatic ground state preparation. Suppose we fix an initial mass $m$ and $\Delta(g)$ to $m=m_0,\Delta(g)=0$, which corresponds to the start of \textcircled{\scriptsize 1}. One can use the initial ground state $\ket{\text{vac}}_0=\ket{0101\cdots01}$ of the Hamiltonian eq.\eqref{eq:initHam_m}. As long as the target mass $m_\text{target}$ and $\Delta(g_\text{target})$ remain in the same phase of the initial state, one can efficiently obtain the value without changing paths. However, if there is a critical point of a 1st order phase transition on the dashed line, one needs to change paths to avoid a 1st order phase transition. For this, we draw an additional path consisting of \textcircled{\scriptsize 2}, \textcircled{\scriptsize 3}, and \textcircled{\scriptsize 4}. In the end of track \textcircled{\scriptsize 2}, one reaches to a point close to $m=0,\Delta(g)<0$, to avoid a strong effect of first order phase transition. Throughout the track \textcircled{\scriptsize 3}, $\Delta(g)$ gradually approaches to the target $\Delta(g_\text{target})$ and the adiabatic process should remain in (or at least close to) the ground state of \textcircled{\scriptsize 2}. 

For quantum simulation, we decompose the time-dependent Hamiltonian~\eqref{eq:Ht} as 
\begin{align}
\begin{aligned}
    H(t)&=H_{\pm}(t)+H_{ZZ}(t)+H_Z(t)\\
    H_{\pm}(t)&=-\frac{{w}(t)}{4a}\sum_{n=1}^{N-1}(X_nX_{n+1}+Y_nY_{n+1})\\
    H_{ZZ}(t)&=\frac{ƒ{\Delta(g)}(t)}{4a}\sum_{n=1}^NZ_nZ_{n+1}\\
    H_Z(t)&=\frac{{m}(t)}{2}\sum_{n=1}^N(-1)^nZ_n+\frac{{\Delta(g)}(t)}{4}\sum_{n=1}^{N-1}(Z_n+Z_{n+1}),
\end{aligned} 
\end{align}
where time-dependent coefficients ${w}(t),{m}(t),{\Delta(g)}(t)$ should be defined so that they agree with a path of computation drawn in Fig.~\ref{fig:path}. The scalar term $\frac{(N-1){\Delta(g)}(t)}{4}$ dose not contribute to time-evolution, thereby we neglect it. For example, in this study we use the following time-dependent parameters:
\begin{align}
    {w}(t)&=
    \begin{cases}
        t/T&\textcircled{\scriptsize 1}\\
        1&\textcircled{\scriptsize 2},\textcircled{\scriptsize 3},\textcircled{\scriptsize 4}
    \end{cases}\\
    {m}(t)&=
    \begin{cases}
        m_0&\textcircled{\scriptsize 1}\\
        m_1\frac{t}{T}+m_0\left(1-\frac{t}{T}\right)&\textcircled{\scriptsize 2}\\
        m_1&\textcircled{\scriptsize 3}\\
        m_2\frac{t}{T}+m_1\left(1-\frac{t}{T}\right)&\textcircled{\scriptsize 4}
    \end{cases}\\
    \Delta(g)(t)&=
    \begin{cases}
        \Delta(g_0)\frac{t^2}{T^2}&\textcircled{\scriptsize 1}\\
        \Delta(g_0)&\textcircled{\scriptsize 2}\\
        \Delta(g_2)\frac{t}{T}+\Delta(g_1)\left(1-\frac{t}{T}\right)&\textcircled{\scriptsize 3}\\
        \Delta(g_2)&\textcircled{\scriptsize 4}
    \end{cases}
\end{align}

Let $T$ be the computational time, $M$ be  a large positive integer (Trotter number), $\delta t=T/M$ be a time-step. We implement the integral~\eqref{eq:integral} in the following way. 
First, the unitary time-evolution for $k$ time steps can be given
\begin{equation}
    U(k\delta t)=e^{-i\delta tH_{\pm}(k\delta t)}e^{-i\delta tH_{ZZ}(k\delta t)}e^{-i\delta tH_{Z}(k\delta t)}
\end{equation}
and the state at $t=k\delta t~(k=1,2,\cdots,T)$ is 
\begin{equation}
\label{eq:ASP_state}
    \ket{\text{vac}(k\delta t)}=U(k\delta t)\cdots U(2\delta t)U(\delta t)\ket{\text{vac}}_0.
\end{equation}
The quantum circuits we implemented are shown in Sec.~\ref{sec:gate}. Therefore at the end of computation $(t=T=M\delta t)$,  the state evolves into 
\begin{equation}
    \ket{\text{vac}}=\prod_{k=1}^Me^{-i\delta tH_{\pm}(k\delta t)}e^{-i\delta tH_{ZZ}(k\delta t)}e^{-i\delta tH_{Z}(k\delta t)}\ket{\text{vac}}_0.
\end{equation}

Throughout this work we chose $N=10,T=100, M=1000,\delta t=T/M=0.1$ and fix the lattice spacing $a$ to 1. 

As a demonstration, we design a path in Fig.~\ref{fig:path}, which consists of four tracks whose initial and target values are given in Table~\ref{tab:parameter}. 

\begin{table}[H]
    \centering
    \begin{tabular}{c|c|c|c|c}\toprule
    & \textcircled{\scriptsize 1}&\textcircled{\scriptsize 2} &\textcircled{\scriptsize 3} &\textcircled{\scriptsize 4}\\ \hline
       initial $(\Delta(g),m)$  & (0,0.7)& (-0.1,0.7) & (-0.1,0)& (-0.6,0)\\
        target $(\Delta(g),m)$ &  (-0.1,0.7)& (-0.1,0) &(-0.6,0) & (-0.6,0.3)\\ \hline
    \end{tabular}
    \caption{Initial and target values for adiabatic state preparation. Each number corresponds to the number in Fig.~\ref{fig:path}, respectively.}
    \label{tab:parameter}
\end{table}

\begin{figure}[H]
    \centering
    \includegraphics[width=\linewidth]{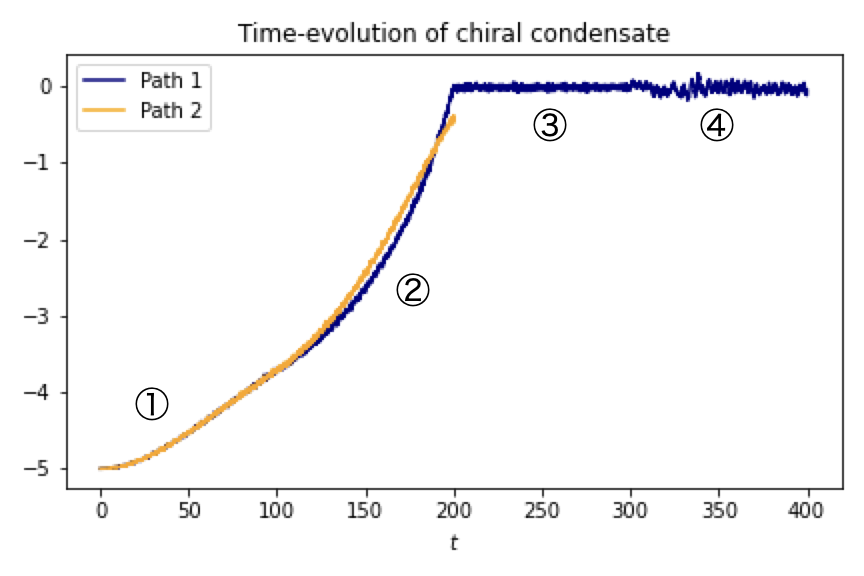}
    \caption{Time-evolution of chiral condensate with the ground state eq.~\eqref{eq:ASP_state} prepared by adiabatic state preparation. Path 1 consists of paths \textcircled{\scriptsize 1} $\cdots$ \textcircled{\scriptsize 4}, and Path 2 is the dashed line in~Fig.~\ref{fig:path}. For each path, the average of 10000 samplings is shown.}
    \label{fig:evolv_CC}
\end{figure}

For comparison, let us consider a linear path connecting $(m_1,\Delta(g_1))$ and $(m_2,\Delta(g_2))$:
\begin{equation}
    \Delta(g)(m)=\frac{\Delta(g_2)-\Delta(g_1)}{m_2-m_1}(m-m_1)+\Delta(g_1). 
\end{equation}
As a function of $t$, $m$ and $\Delta(g)$ change as follows
\begin{align}
\begin{aligned}
    m(t)&=m_1\left(1-\frac{t}{T}\right)+m_2\frac{t}{T}\\
    \Delta(g)(t)&=\Delta(g_1)\left(1-\frac{t}{T}\right)+\Delta(g_2)\frac{t}{T}
\end{aligned}    
\end{align}
The results are shown in Fig.~\ref{fig:evolv_CC}. It is clear that results of Path 1 is consistent with Fig.~\ref{fig:chiral}. Path 2 continues to take the same values as Path 1 when there is no phase transition, while after crossing the phase transition point it deviates significantly from the exact theoretical values. Note that the time steps for Path 1 and Path 2 are the same.

\section*{References}
\bibliographystyle{apsrev4-1.bst}
\bibliography{ref}

%merlin.mbs apsrev4-1.bst 2010-07-25 4.21a (PWD, AO, DPC) hacked
%Control: key (0)
%Control: author (72) initials jnrlst
%Control: editor formatted (1) identically to author
%Control: production of article title (-1) disabled
%Control: page (0) single
%Control: year (1) truncated
%Control: production of eprint (0) enabled
\begin{thebibliography}{58}%
\makeatletter
\providecommand \@ifxundefined [1]{%
 \@ifx{#1\undefined}
}%
\providecommand \@ifnum [1]{%
 \ifnum #1\expandafter \@firstoftwo
 \else \expandafter \@secondoftwo
 \fi
}%
\providecommand \@ifx [1]{%
 \ifx #1\expandafter \@firstoftwo
 \else \expandafter \@secondoftwo
 \fi
}%
\providecommand \natexlab [1]{#1}%
\providecommand \enquote  [1]{``#1''}%
\providecommand \bibnamefont  [1]{#1}%
\providecommand \bibfnamefont [1]{#1}%
\providecommand \citenamefont [1]{#1}%
\providecommand \href@noop [0]{\@secondoftwo}%
\providecommand \href [0]{\begingroup \@sanitize@url \@href}%
\providecommand \@href[1]{\@@startlink{#1}\@@href}%
\providecommand \@@href[1]{\endgroup#1\@@endlink}%
\providecommand \@sanitize@url [0]{\catcode `\\12\catcode `\$12\catcode
  `\&12\catcode `\#12\catcode `\^12\catcode `\_12\catcode `\%12\relax}%
\providecommand \@@startlink[1]{}%
\providecommand \@@endlink[0]{}%
\providecommand \url  [0]{\begingroup\@sanitize@url \@url }%
\providecommand \@url [1]{\endgroup\@href {#1}{\urlprefix }}%
\providecommand \urlprefix  [0]{URL }%
\providecommand \Eprint [0]{\href }%
\providecommand \doibase [0]{http://dx.doi.org/}%
\providecommand \selectlanguage [0]{\@gobble}%
\providecommand \bibinfo  [0]{\@secondoftwo}%
\providecommand \bibfield  [0]{\@secondoftwo}%
\providecommand \translation [1]{[#1]}%
\providecommand \BibitemOpen [0]{}%
\providecommand \bibitemStop [0]{}%
\providecommand \bibitemNoStop [0]{.\EOS\space}%
\providecommand \EOS [0]{\spacefactor3000\relax}%
\providecommand \BibitemShut  [1]{\csname bibitem#1\endcsname}%
\let\auto@bib@innerbib\@empty
%</preamble>
\bibitem [{\citenamefont {{Ryu}}\ and\ \citenamefont
  {{Takayanagi}}(2006)}]{2006PhRvL..96r1602R}%
  \BibitemOpen
  \bibfield  {author} {\bibinfo {author} {\bibfnamefont {S.}~\bibnamefont
  {{Ryu}}}\ and\ \bibinfo {author} {\bibfnamefont {T.}~\bibnamefont
  {{Takayanagi}}},\ }\href {\doibase 10.1103/PhysRevLett.96.181602} {\bibfield
  {journal} {\bibinfo  {journal} {Phys. Rev. Lett.}\ }\textbf {\bibinfo
  {volume} {96}},\ \bibinfo {eid} {181602} (\bibinfo {year} {2006})},\ \Eprint
  {http://arxiv.org/abs/hep-th/0603001} {arXiv:hep-th/0603001 [hep-th]}
  \BibitemShut {NoStop}%
\bibitem [{\citenamefont {{Yoshida}}(2019)}]{2019JHEP...10..132Y}%
  \BibitemOpen
  \bibfield  {author} {\bibinfo {author} {\bibfnamefont {B.}~\bibnamefont
  {{Yoshida}}},\ }\href {\doibase 10.1007/JHEP10(2019)132} {\bibfield
  {journal} {\bibinfo  {journal} {Journal of High Energy Physics}\ }\textbf
  {\bibinfo {volume} {2019}},\ \bibinfo {eid} {132} (\bibinfo {year} {2019})},\
  \Eprint {http://arxiv.org/abs/1902.09763} {arXiv:1902.09763 [hep-th]}
  \BibitemShut {NoStop}%
\bibitem [{\citenamefont {{Hayden}}\ and\ \citenamefont
  {{Preskill}}(2007)}]{2007JHEP...09..120H}%
  \BibitemOpen
  \bibfield  {author} {\bibinfo {author} {\bibfnamefont {P.}~\bibnamefont
  {{Hayden}}}\ and\ \bibinfo {author} {\bibfnamefont {J.}~\bibnamefont
  {{Preskill}}},\ }\href {\doibase 10.1088/1126-6708/2007/09/120} {\bibfield
  {journal} {\bibinfo  {journal} {Journal of High Energy Physics}\ }\textbf
  {\bibinfo {volume} {2007}},\ \bibinfo {eid} {120} (\bibinfo {year} {2007})},\
  \Eprint {http://arxiv.org/abs/0708.4025} {arXiv:0708.4025 [hep-th]}
  \BibitemShut {NoStop}%
\bibitem [{\citenamefont {Minato}\ \emph {et~al.}(2022)\citenamefont {Minato},
  \citenamefont {Sugimoto}, \citenamefont {Kuwahara},\ and\ \citenamefont
  {Saito}}]{PhysRevLett.128.010603}%
  \BibitemOpen
  \bibfield  {author} {\bibinfo {author} {\bibfnamefont {T.}~\bibnamefont
  {Minato}}, \bibinfo {author} {\bibfnamefont {K.}~\bibnamefont {Sugimoto}},
  \bibinfo {author} {\bibfnamefont {T.}~\bibnamefont {Kuwahara}}, \ and\
  \bibinfo {author} {\bibfnamefont {K.}~\bibnamefont {Saito}},\ }\href
  {\doibase 10.1103/PhysRevLett.128.010603} {\bibfield  {journal} {\bibinfo
  {journal} {Phys. Rev. Lett.}\ }\textbf {\bibinfo {volume} {128}},\ \bibinfo
  {pages} {010603} (\bibinfo {year} {2022})}\BibitemShut {NoStop}%
\bibitem [{\citenamefont {Hashizume}\ \emph {et~al.}(2022)\citenamefont
  {Hashizume}, \citenamefont {Bentsen},\ and\ \citenamefont
  {Daley}}]{PhysRevResearch.4.013174}%
  \BibitemOpen
  \bibfield  {author} {\bibinfo {author} {\bibfnamefont {T.}~\bibnamefont
  {Hashizume}}, \bibinfo {author} {\bibfnamefont {G.}~\bibnamefont {Bentsen}},
  \ and\ \bibinfo {author} {\bibfnamefont {A.~J.}\ \bibnamefont {Daley}},\
  }\href {\doibase 10.1103/PhysRevResearch.4.013174} {\bibfield  {journal}
  {\bibinfo  {journal} {Phys. Rev. Res.}\ }\textbf {\bibinfo {volume} {4}},\
  \bibinfo {pages} {013174} (\bibinfo {year} {2022})}\BibitemShut {NoStop}%
\bibitem [{\citenamefont {Briegel}\ \emph {et~al.}(2009)\citenamefont
  {Briegel}, \citenamefont {Browne}, \citenamefont {D{\"u}r}, \citenamefont
  {Raussendorf},\ and\ \citenamefont {Van~den Nest}}]{briegel2009measurement}%
  \BibitemOpen
  \bibfield  {author} {\bibinfo {author} {\bibfnamefont {H.~J.}\ \bibnamefont
  {Briegel}}, \bibinfo {author} {\bibfnamefont {D.~E.}\ \bibnamefont {Browne}},
  \bibinfo {author} {\bibfnamefont {W.}~\bibnamefont {D{\"u}r}}, \bibinfo
  {author} {\bibfnamefont {R.}~\bibnamefont {Raussendorf}}, \ and\ \bibinfo
  {author} {\bibfnamefont {M.}~\bibnamefont {Van~den Nest}},\ }\href@noop {}
  {\bibfield  {journal} {\bibinfo  {journal} {Nature Physics}\ }\textbf
  {\bibinfo {volume} {5}},\ \bibinfo {pages} {19} (\bibinfo {year}
  {2009})}\BibitemShut {NoStop}%
\bibitem [{\citenamefont {{Gebhart}}\ \emph {et~al.}(2020)\citenamefont
  {{Gebhart}}, \citenamefont {{Snizhko}}, \citenamefont {{Wellens}},
  \citenamefont {{Buchleitner}}, \citenamefont {{Romito}},\ and\ \citenamefont
  {{Gefen}}}]{2020PNAS..117.5706G}%
  \BibitemOpen
  \bibfield  {author} {\bibinfo {author} {\bibfnamefont {V.}~\bibnamefont
  {{Gebhart}}}, \bibinfo {author} {\bibfnamefont {K.}~\bibnamefont
  {{Snizhko}}}, \bibinfo {author} {\bibfnamefont {T.}~\bibnamefont
  {{Wellens}}}, \bibinfo {author} {\bibfnamefont {A.}~\bibnamefont
  {{Buchleitner}}}, \bibinfo {author} {\bibfnamefont {A.}~\bibnamefont
  {{Romito}}}, \ and\ \bibinfo {author} {\bibfnamefont {Y.}~\bibnamefont
  {{Gefen}}},\ }\href {\doibase 10.1073/pnas.1911620117} {\bibfield  {journal}
  {\bibinfo  {journal} {Proceedings of the National Academy of Science}\
  }\textbf {\bibinfo {volume} {117}},\ \bibinfo {pages} {5706} (\bibinfo {year}
  {2020})},\ \Eprint {http://arxiv.org/abs/1905.01147} {arXiv:1905.01147
  [quant-ph]} \BibitemShut {NoStop}%
\bibitem [{\citenamefont {Stephen}\ \emph {et~al.}(2022)\citenamefont
  {Stephen}, \citenamefont {Ho}, \citenamefont {Wei}, \citenamefont
  {Raussendorf},\ and\ \citenamefont {Verresen}}]{stephen2022universal}%
  \BibitemOpen
  \bibfield  {author} {\bibinfo {author} {\bibfnamefont {D.~T.}\ \bibnamefont
  {Stephen}}, \bibinfo {author} {\bibfnamefont {W.~W.}\ \bibnamefont {Ho}},
  \bibinfo {author} {\bibfnamefont {T.-C.}\ \bibnamefont {Wei}}, \bibinfo
  {author} {\bibfnamefont {R.}~\bibnamefont {Raussendorf}}, \ and\ \bibinfo
  {author} {\bibfnamefont {R.}~\bibnamefont {Verresen}},\ }\href@noop {}
  {\bibfield  {journal} {\bibinfo  {journal} {arXiv preprint arXiv:2209.06191}\
  } (\bibinfo {year} {2022})}\BibitemShut {NoStop}%
\bibitem [{\citenamefont {{Koh}}\ \emph {et~al.}(2022)\citenamefont {{Koh}},
  \citenamefont {{Sun}}, \citenamefont {{Motta}},\ and\ \citenamefont
  {{Minnich}}}]{2022arXiv220304338K}%
  \BibitemOpen
  \bibfield  {author} {\bibinfo {author} {\bibfnamefont {J.~M.}\ \bibnamefont
  {{Koh}}}, \bibinfo {author} {\bibfnamefont {S.-N.}\ \bibnamefont {{Sun}}},
  \bibinfo {author} {\bibfnamefont {M.}~\bibnamefont {{Motta}}}, \ and\
  \bibinfo {author} {\bibfnamefont {A.~J.}\ \bibnamefont {{Minnich}}},\ }\href
  {\doibase 10.48550/arXiv.2203.04338} {\bibfield  {journal} {\bibinfo
  {journal} {arXiv e-prints}\ ,\ \bibinfo {eid} {arXiv:2203.04338}} (\bibinfo
  {year} {2022})},\ \Eprint {http://arxiv.org/abs/2203.04338} {arXiv:2203.04338
  [quant-ph]} \BibitemShut {NoStop}%
\bibitem [{\citenamefont {{Torres}}\ \emph {et~al.}(2023)\citenamefont
  {{Torres}}, \citenamefont {{Wurtz}}, \citenamefont {{Polo-G{\'o}mez}},\ and\
  \citenamefont {{Mart{\'\i}n-Mart{\'\i}nez}}}]{2023arXiv230108775T}%
  \BibitemOpen
  \bibfield  {author} {\bibinfo {author} {\bibfnamefont {B.~d. S.~L.}\
  \bibnamefont {{Torres}}}, \bibinfo {author} {\bibfnamefont {K.}~\bibnamefont
  {{Wurtz}}}, \bibinfo {author} {\bibfnamefont {J.}~\bibnamefont
  {{Polo-G{\'o}mez}}}, \ and\ \bibinfo {author} {\bibfnamefont
  {E.}~\bibnamefont {{Mart{\'\i}n-Mart{\'\i}nez}}},\ }\href {\doibase
  10.48550/arXiv.2301.08775} {\bibfield  {journal} {\bibinfo  {journal} {arXiv
  e-prints}\ ,\ \bibinfo {eid} {arXiv:2301.08775}} (\bibinfo {year} {2023})},\
  \Eprint {http://arxiv.org/abs/2301.08775} {arXiv:2301.08775 [quant-ph]}
  \BibitemShut {NoStop}%
\bibitem [{\citenamefont {{Maeso-Garc{\'\i}a}}\ \emph
  {et~al.}(2022)\citenamefont {{Maeso-Garc{\'\i}a}}, \citenamefont
  {{Polo-G{\'o}mez}},\ and\ \citenamefont
  {{Mart{\'\i}n-Mart{\'\i}nez}}}]{2022arXiv221005692M}%
  \BibitemOpen
  \bibfield  {author} {\bibinfo {author} {\bibfnamefont {H.}~\bibnamefont
  {{Maeso-Garc{\'\i}a}}}, \bibinfo {author} {\bibfnamefont {J.}~\bibnamefont
  {{Polo-G{\'o}mez}}}, \ and\ \bibinfo {author} {\bibfnamefont
  {E.}~\bibnamefont {{Mart{\'\i}n-Mart{\'\i}nez}}},\ }\href {\doibase
  10.48550/arXiv.2210.05692} {\bibfield  {journal} {\bibinfo  {journal} {arXiv
  e-prints}\ ,\ \bibinfo {eid} {arXiv:2210.05692}} (\bibinfo {year} {2022})},\
  \Eprint {http://arxiv.org/abs/2210.05692} {arXiv:2210.05692 [quant-ph]}
  \BibitemShut {NoStop}%
\bibitem [{\citenamefont {Perche}\ and\ \citenamefont
  {Mart\'{\i}n-Mart\'{\i}nez}(2022)}]{PhysRevD.105.066011}%
  \BibitemOpen
  \bibfield  {author} {\bibinfo {author} {\bibfnamefont {T.~R.}\ \bibnamefont
  {Perche}}\ and\ \bibinfo {author} {\bibfnamefont {E.}~\bibnamefont
  {Mart\'{\i}n-Mart\'{\i}nez}},\ }\href {\doibase 10.1103/PhysRevD.105.066011}
  {\bibfield  {journal} {\bibinfo  {journal} {Phys. Rev. D}\ }\textbf {\bibinfo
  {volume} {105}},\ \bibinfo {pages} {066011} (\bibinfo {year}
  {2022})}\BibitemShut {NoStop}%
\bibitem [{\citenamefont {{Rodr{\'\i}guez-Briones}}\ \emph
  {et~al.}(2022)\citenamefont {{Rodr{\'\i}guez-Briones}}, \citenamefont
  {{Katiyar}}, \citenamefont {{Laflamme}},\ and\ \citenamefont
  {{Mart{\'\i}n-Mart{\'\i}nez}}}]{2022arXiv220316269R}%
  \BibitemOpen
  \bibfield  {author} {\bibinfo {author} {\bibfnamefont {N.~A.}\ \bibnamefont
  {{Rodr{\'\i}guez-Briones}}}, \bibinfo {author} {\bibfnamefont
  {H.}~\bibnamefont {{Katiyar}}}, \bibinfo {author} {\bibfnamefont
  {R.}~\bibnamefont {{Laflamme}}}, \ and\ \bibinfo {author} {\bibfnamefont
  {E.}~\bibnamefont {{Mart{\'\i}n-Mart{\'\i}nez}}},\ }\href {\doibase
  10.48550/arXiv.2203.16269} {\bibfield  {journal} {\bibinfo  {journal} {arXiv
  e-prints}\ ,\ \bibinfo {eid} {arXiv:2203.16269}} (\bibinfo {year} {2022})},\
  \Eprint {http://arxiv.org/abs/2203.16269} {arXiv:2203.16269 [quant-ph]}
  \BibitemShut {NoStop}%
\bibitem [{\citenamefont {Polo-G\'omez}\ \emph {et~al.}(2022)\citenamefont
  {Polo-G\'omez}, \citenamefont {Garay},\ and\ \citenamefont
  {Mart\'{\i}n-Mart\'{\i}nez}}]{PhysRevD.105.065003}%
  \BibitemOpen
  \bibfield  {author} {\bibinfo {author} {\bibfnamefont {J.}~\bibnamefont
  {Polo-G\'omez}}, \bibinfo {author} {\bibfnamefont {L.~J.}\ \bibnamefont
  {Garay}}, \ and\ \bibinfo {author} {\bibfnamefont {E.}~\bibnamefont
  {Mart\'{\i}n-Mart\'{\i}nez}},\ }\href {\doibase 10.1103/PhysRevD.105.065003}
  {\bibfield  {journal} {\bibinfo  {journal} {Phys. Rev. D}\ }\textbf {\bibinfo
  {volume} {105}},\ \bibinfo {pages} {065003} (\bibinfo {year}
  {2022})}\BibitemShut {NoStop}%
\bibitem [{\citenamefont {Hotta}(2008{\natexlab{a}})}]{HOTTA20085671}%
  \BibitemOpen
  \bibfield  {author} {\bibinfo {author} {\bibfnamefont {M.}~\bibnamefont
  {Hotta}},\ }\href {\doibase https://doi.org/10.1016/j.physleta.2008.07.007}
  {\bibfield  {journal} {\bibinfo  {journal} {Physics Letters A}\ }\textbf
  {\bibinfo {volume} {372}},\ \bibinfo {pages} {5671} (\bibinfo {year}
  {2008}{\natexlab{a}})}\BibitemShut {NoStop}%
\bibitem [{\citenamefont {{Hotta}}(2009{\natexlab{a}})}]{2009JPSJ...78c4001H}%
  \BibitemOpen
  \bibfield  {author} {\bibinfo {author} {\bibfnamefont {M.}~\bibnamefont
  {{Hotta}}},\ }\href {\doibase 10.1143/JPSJ.78.034001} {\bibfield  {journal}
  {\bibinfo  {journal} {Journal of the Physical Society of Japan}\ }\textbf
  {\bibinfo {volume} {78}},\ \bibinfo {pages} {034001} (\bibinfo {year}
  {2009}{\natexlab{a}})},\ \Eprint {http://arxiv.org/abs/0803.0348}
  {arXiv:0803.0348 [quant-ph]} \BibitemShut {NoStop}%
\bibitem [{\citenamefont {{Trevison}}\ and\ \citenamefont
  {{Hotta}}(2015)}]{2015JPhA...48q5302T}%
  \BibitemOpen
  \bibfield  {author} {\bibinfo {author} {\bibfnamefont {J.}~\bibnamefont
  {{Trevison}}}\ and\ \bibinfo {author} {\bibfnamefont {M.}~\bibnamefont
  {{Hotta}}},\ }\href {\doibase 10.1088/1751-8113/48/17/175302} {\bibfield
  {journal} {\bibinfo  {journal} {Journal of Physics A Mathematical General}\
  }\textbf {\bibinfo {volume} {48}},\ \bibinfo {eid} {175302} (\bibinfo {year}
  {2015})},\ \Eprint {http://arxiv.org/abs/1411.7495} {arXiv:1411.7495
  [quant-ph]} \BibitemShut {NoStop}%
\bibitem [{\citenamefont {{Hotta}}(2009{\natexlab{b}})}]{2009PhRvA..80d2323H}%
  \BibitemOpen
  \bibfield  {author} {\bibinfo {author} {\bibfnamefont {M.}~\bibnamefont
  {{Hotta}}},\ }\href {\doibase 10.1103/PhysRevA.80.042323} {\bibfield
  {journal} {\bibinfo  {journal} {Phys. Rev. A}\ }\textbf {\bibinfo {volume}
  {80}},\ \bibinfo {eid} {042323} (\bibinfo {year} {2009}{\natexlab{b}})},\
  \Eprint {http://arxiv.org/abs/0908.2824} {arXiv:0908.2824 [quant-ph]}
  \BibitemShut {NoStop}%
\bibitem [{\citenamefont {Nambu}\ and\ \citenamefont
  {Hotta}(2010)}]{PhysRevA.82.042329}%
  \BibitemOpen
  \bibfield  {author} {\bibinfo {author} {\bibfnamefont {Y.}~\bibnamefont
  {Nambu}}\ and\ \bibinfo {author} {\bibfnamefont {M.}~\bibnamefont {Hotta}},\
  }\href {\doibase 10.1103/PhysRevA.82.042329} {\bibfield  {journal} {\bibinfo
  {journal} {Phys. Rev. A}\ }\textbf {\bibinfo {volume} {82}},\ \bibinfo
  {pages} {042329} (\bibinfo {year} {2010})}\BibitemShut {NoStop}%
\bibitem [{\citenamefont {Hotta}(2010)}]{Hotta_2010}%
  \BibitemOpen
  \bibfield  {author} {\bibinfo {author} {\bibfnamefont {M.}~\bibnamefont
  {Hotta}},\ }\href {\doibase 10.1088/1751-8113/43/10/105305} {\bibfield
  {journal} {\bibinfo  {journal} {Journal of Physics A: Mathematical and
  Theoretical}\ }\textbf {\bibinfo {volume} {43}},\ \bibinfo {pages} {105305}
  (\bibinfo {year} {2010})}\BibitemShut {NoStop}%
\bibitem [{\citenamefont {{Ikeda}}(2023)}]{2023arXiv230102666I}%
  \BibitemOpen
  \bibfield  {author} {\bibinfo {author} {\bibfnamefont {K.}~\bibnamefont
  {{Ikeda}}},\ }\href@noop {} {\bibfield  {journal} {\bibinfo  {journal} {arXiv
  e-prints}\ ,\ \bibinfo {eid} {arXiv:2301.02666}} (\bibinfo {year} {2023})},\
  \Eprint {http://arxiv.org/abs/2301.02666} {arXiv:2301.02666 [quant-ph]}
  \BibitemShut {NoStop}%
\bibitem [{\citenamefont {Bennett}\ \emph {et~al.}(1993)\citenamefont
  {Bennett}, \citenamefont {Brassard}, \citenamefont {Cr\'epeau}, \citenamefont
  {Jozsa}, \citenamefont {Peres},\ and\ \citenamefont
  {Wootters}}]{PhysRevLett.70.1895}%
  \BibitemOpen
  \bibfield  {author} {\bibinfo {author} {\bibfnamefont {C.~H.}\ \bibnamefont
  {Bennett}}, \bibinfo {author} {\bibfnamefont {G.}~\bibnamefont {Brassard}},
  \bibinfo {author} {\bibfnamefont {C.}~\bibnamefont {Cr\'epeau}}, \bibinfo
  {author} {\bibfnamefont {R.}~\bibnamefont {Jozsa}}, \bibinfo {author}
  {\bibfnamefont {A.}~\bibnamefont {Peres}}, \ and\ \bibinfo {author}
  {\bibfnamefont {W.~K.}\ \bibnamefont {Wootters}},\ }\href {\doibase
  10.1103/PhysRevLett.70.1895} {\bibfield  {journal} {\bibinfo  {journal}
  {Phys. Rev. Lett.}\ }\textbf {\bibinfo {volume} {70}},\ \bibinfo {pages}
  {1895} (\bibinfo {year} {1993})}\BibitemShut {NoStop}%
\bibitem [{\citenamefont {Furusawa}\ \emph {et~al.}(1998)\citenamefont
  {Furusawa}, \citenamefont {S{\o}rensen}, \citenamefont {Braunstein},
  \citenamefont {Fuchs}, \citenamefont {Kimble},\ and\ \citenamefont
  {Polzik}}]{furusawa1998unconditional}%
  \BibitemOpen
  \bibfield  {author} {\bibinfo {author} {\bibfnamefont {A.}~\bibnamefont
  {Furusawa}}, \bibinfo {author} {\bibfnamefont {J.~L.}\ \bibnamefont
  {S{\o}rensen}}, \bibinfo {author} {\bibfnamefont {S.~L.}\ \bibnamefont
  {Braunstein}}, \bibinfo {author} {\bibfnamefont {C.~A.}\ \bibnamefont
  {Fuchs}}, \bibinfo {author} {\bibfnamefont {H.~J.}\ \bibnamefont {Kimble}}, \
  and\ \bibinfo {author} {\bibfnamefont {E.~S.}\ \bibnamefont {Polzik}},\
  }\href@noop {} {\bibfield  {journal} {\bibinfo  {journal} {science}\ }\textbf
  {\bibinfo {volume} {282}},\ \bibinfo {pages} {706} (\bibinfo {year}
  {1998})}\BibitemShut {NoStop}%
\bibitem [{\citenamefont {{Pirandola}}\ \emph {et~al.}(2015)\citenamefont
  {{Pirandola}}, \citenamefont {{Eisert}}, \citenamefont {{Weedbrook}},
  \citenamefont {{Furusawa}},\ and\ \citenamefont
  {{Braunstein}}}]{2015NaPho...9..641P}%
  \BibitemOpen
  \bibfield  {author} {\bibinfo {author} {\bibfnamefont {S.}~\bibnamefont
  {{Pirandola}}}, \bibinfo {author} {\bibfnamefont {J.}~\bibnamefont
  {{Eisert}}}, \bibinfo {author} {\bibfnamefont {C.}~\bibnamefont
  {{Weedbrook}}}, \bibinfo {author} {\bibfnamefont {A.}~\bibnamefont
  {{Furusawa}}}, \ and\ \bibinfo {author} {\bibfnamefont {S.~L.}\ \bibnamefont
  {{Braunstein}}},\ }\href {\doibase 10.1038/nphoton.2015.154} {\bibfield
  {journal} {\bibinfo  {journal} {Nature Photonics}\ }\textbf {\bibinfo
  {volume} {9}},\ \bibinfo {pages} {641} (\bibinfo {year} {2015})},\ \Eprint
  {http://arxiv.org/abs/1505.07831} {arXiv:1505.07831 [quant-ph]} \BibitemShut
  {NoStop}%
\bibitem [{\citenamefont {Takeda}\ \emph {et~al.}(2013)\citenamefont {Takeda},
  \citenamefont {Mizuta}, \citenamefont {Fuwa}, \citenamefont {Van~Loock},\
  and\ \citenamefont {Furusawa}}]{takeda2013deterministic}%
  \BibitemOpen
  \bibfield  {author} {\bibinfo {author} {\bibfnamefont {S.}~\bibnamefont
  {Takeda}}, \bibinfo {author} {\bibfnamefont {T.}~\bibnamefont {Mizuta}},
  \bibinfo {author} {\bibfnamefont {M.}~\bibnamefont {Fuwa}}, \bibinfo {author}
  {\bibfnamefont {P.}~\bibnamefont {Van~Loock}}, \ and\ \bibinfo {author}
  {\bibfnamefont {A.}~\bibnamefont {Furusawa}},\ }\href@noop {} {\bibfield
  {journal} {\bibinfo  {journal} {Nature}\ }\textbf {\bibinfo {volume} {500}},\
  \bibinfo {pages} {315} (\bibinfo {year} {2013})}\BibitemShut {NoStop}%
\bibitem [{\citenamefont {Thotakura}\ and\ \citenamefont
  {Wei}(2022)}]{thotakura2022quantum}%
  \BibitemOpen
  \bibfield  {author} {\bibinfo {author} {\bibfnamefont {B.}~\bibnamefont
  {Thotakura}}\ and\ \bibinfo {author} {\bibfnamefont {T.-C.}\ \bibnamefont
  {Wei}},\ }\href@noop {} {\bibfield  {journal} {\bibinfo  {journal} {arXiv
  preprint arXiv:2209.07021}\ } (\bibinfo {year} {2022})}\BibitemShut {NoStop}%
\bibitem [{\citenamefont {Izergin}\ and\ \citenamefont
  {Korepin}(1982)}]{IZERGIN1982401}%
  \BibitemOpen
  \bibfield  {author} {\bibinfo {author} {\bibfnamefont {A.}~\bibnamefont
  {Izergin}}\ and\ \bibinfo {author} {\bibfnamefont {V.}~\bibnamefont
  {Korepin}},\ }\href {\doibase https://doi.org/10.1016/0550-3213(82)90365-0}
  {\bibfield  {journal} {\bibinfo  {journal} {Nuclear Physics B}\ }\textbf
  {\bibinfo {volume} {205}},\ \bibinfo {pages} {401} (\bibinfo {year}
  {1982})}\BibitemShut {NoStop}%
\bibitem [{\citenamefont {Ikeda}\ \emph {et~al.}(2021)\citenamefont {Ikeda},
  \citenamefont {Kharzeev},\ and\ \citenamefont
  {Kikuchi}}]{PhysRevD.103.L071502}%
  \BibitemOpen
  \bibfield  {author} {\bibinfo {author} {\bibfnamefont {K.}~\bibnamefont
  {Ikeda}}, \bibinfo {author} {\bibfnamefont {D.~E.}\ \bibnamefont {Kharzeev}},
  \ and\ \bibinfo {author} {\bibfnamefont {Y.}~\bibnamefont {Kikuchi}},\ }\href
  {\doibase 10.1103/PhysRevD.103.L071502} {\bibfield  {journal} {\bibinfo
  {journal} {Phys. Rev. D}\ }\textbf {\bibinfo {volume} {103}},\ \bibinfo
  {pages} {L071502} (\bibinfo {year} {2021})}\BibitemShut {NoStop}%
\bibitem [{\citenamefont {Honda}\ \emph {et~al.}(2022)\citenamefont {Honda},
  \citenamefont {Itou}, \citenamefont {Kikuchi},\ and\ \citenamefont
  {Tanizaki}}]{10.1093/ptep/ptac007}%
  \BibitemOpen
  \bibfield  {author} {\bibinfo {author} {\bibfnamefont {M.}~\bibnamefont
  {Honda}}, \bibinfo {author} {\bibfnamefont {E.}~\bibnamefont {Itou}},
  \bibinfo {author} {\bibfnamefont {Y.}~\bibnamefont {Kikuchi}}, \ and\
  \bibinfo {author} {\bibfnamefont {Y.}~\bibnamefont {Tanizaki}},\ }\href
  {\doibase 10.1093/ptep/ptac007} {\bibfield  {journal} {\bibinfo  {journal}
  {Progress of Theoretical and Experimental Physics}\ }\textbf {\bibinfo
  {volume} {2022}} (\bibinfo {year} {2022}),\ 10.1093/ptep/ptac007},\ \bibinfo
  {note} {033B01},\ \Eprint
  {http://arxiv.org/abs/https://academic.oup.com/ptep/article-pdf/2022/3/033B01/42782471/ptac007.pdf}
  {https://academic.oup.com/ptep/article-pdf/2022/3/033B01/42782471/ptac007.pdf}
  \BibitemShut {NoStop}%
\bibitem [{\citenamefont {Kharzeev}\ and\ \citenamefont
  {Kikuchi}(2020)}]{PhysRevResearch.2.023342}%
  \BibitemOpen
  \bibfield  {author} {\bibinfo {author} {\bibfnamefont {D.~E.}\ \bibnamefont
  {Kharzeev}}\ and\ \bibinfo {author} {\bibfnamefont {Y.}~\bibnamefont
  {Kikuchi}},\ }\href {\doibase 10.1103/PhysRevResearch.2.023342} {\bibfield
  {journal} {\bibinfo  {journal} {Phys. Rev. Res.}\ }\textbf {\bibinfo {volume}
  {2}},\ \bibinfo {pages} {023342} (\bibinfo {year} {2020})}\BibitemShut
  {NoStop}%
\bibitem [{\citenamefont {{Chakraborty}}\ \emph {et~al.}(2020)\citenamefont
  {{Chakraborty}}, \citenamefont {{Honda}}, \citenamefont {{Izubuchi}},
  \citenamefont {{Kikuchi}},\ and\ \citenamefont
  {{Tomiya}}}]{2020arXiv200100485C}%
  \BibitemOpen
  \bibfield  {author} {\bibinfo {author} {\bibfnamefont {B.}~\bibnamefont
  {{Chakraborty}}}, \bibinfo {author} {\bibfnamefont {M.}~\bibnamefont
  {{Honda}}}, \bibinfo {author} {\bibfnamefont {T.}~\bibnamefont {{Izubuchi}}},
  \bibinfo {author} {\bibfnamefont {Y.}~\bibnamefont {{Kikuchi}}}, \ and\
  \bibinfo {author} {\bibfnamefont {A.}~\bibnamefont {{Tomiya}}},\ }\href
  {\doibase 10.48550/arXiv.2001.00485} {\bibfield  {journal} {\bibinfo
  {journal} {arXiv e-prints}\ ,\ \bibinfo {eid} {arXiv:2001.00485}} (\bibinfo
  {year} {2020})},\ \Eprint {http://arxiv.org/abs/2001.00485} {arXiv:2001.00485
  [hep-lat]} \BibitemShut {NoStop}%
\bibitem [{\citenamefont {Mishra}\ \emph {et~al.}(2020)\citenamefont {Mishra},
  \citenamefont {Thompson}, \citenamefont {Pooser},\ and\ \citenamefont
  {Siopsis}}]{Mishra_2020}%
  \BibitemOpen
  \bibfield  {author} {\bibinfo {author} {\bibfnamefont {C.}~\bibnamefont
  {Mishra}}, \bibinfo {author} {\bibfnamefont {S.}~\bibnamefont {Thompson}},
  \bibinfo {author} {\bibfnamefont {R.}~\bibnamefont {Pooser}}, \ and\ \bibinfo
  {author} {\bibfnamefont {G.}~\bibnamefont {Siopsis}},\ }\href {\doibase
  10.1088/2058-9565/ab8f63} {\bibfield  {journal} {\bibinfo  {journal} {Quantum
  Science and Technology}\ }\textbf {\bibinfo {volume} {5}},\ \bibinfo {pages}
  {035010} (\bibinfo {year} {2020})}\BibitemShut {NoStop}%
\bibitem [{\citenamefont {Klco}\ \emph {et~al.}(2018)\citenamefont {Klco},
  \citenamefont {Dumitrescu}, \citenamefont {McCaskey}, \citenamefont {Morris},
  \citenamefont {Pooser}, \citenamefont {Sanz}, \citenamefont {Solano},
  \citenamefont {Lougovski},\ and\ \citenamefont {Savage}}]{klco2018quantum}%
  \BibitemOpen
  \bibfield  {author} {\bibinfo {author} {\bibfnamefont {N.}~\bibnamefont
  {Klco}}, \bibinfo {author} {\bibfnamefont {E.~F.}\ \bibnamefont
  {Dumitrescu}}, \bibinfo {author} {\bibfnamefont {A.~J.}\ \bibnamefont
  {McCaskey}}, \bibinfo {author} {\bibfnamefont {T.~D.}\ \bibnamefont
  {Morris}}, \bibinfo {author} {\bibfnamefont {R.~C.}\ \bibnamefont {Pooser}},
  \bibinfo {author} {\bibfnamefont {M.}~\bibnamefont {Sanz}}, \bibinfo {author}
  {\bibfnamefont {E.}~\bibnamefont {Solano}}, \bibinfo {author} {\bibfnamefont
  {P.}~\bibnamefont {Lougovski}}, \ and\ \bibinfo {author} {\bibfnamefont
  {M.~J.}\ \bibnamefont {Savage}},\ }\href@noop {} {\bibfield  {journal}
  {\bibinfo  {journal} {Physical Review A}\ }\textbf {\bibinfo {volume} {98}},\
  \bibinfo {pages} {032331} (\bibinfo {year} {2018})}\BibitemShut {NoStop}%
\bibitem [{\citenamefont {Thirring}(1958)}]{THIRRING195891}%
  \BibitemOpen
  \bibfield  {author} {\bibinfo {author} {\bibfnamefont {W.~E.}\ \bibnamefont
  {Thirring}},\ }\href {\doibase https://doi.org/10.1016/0003-4916(58)90015-0}
  {\bibfield  {journal} {\bibinfo  {journal} {Annals of Physics}\ }\textbf
  {\bibinfo {volume} {3}},\ \bibinfo {pages} {91} (\bibinfo {year}
  {1958})}\BibitemShut {NoStop}%
\bibitem [{\citenamefont {{Faber}}\ and\ \citenamefont
  {{Ivanov}}(2001)}]{2001EPJC...20..723F}%
  \BibitemOpen
  \bibfield  {author} {\bibinfo {author} {\bibfnamefont {M.}~\bibnamefont
  {{Faber}}}\ and\ \bibinfo {author} {\bibfnamefont {A.~N.}\ \bibnamefont
  {{Ivanov}}},\ }\href {\doibase 10.1007/s100520100694} {\bibfield  {journal}
  {\bibinfo  {journal} {European Physical Journal C}\ }\textbf {\bibinfo
  {volume} {20}},\ \bibinfo {pages} {723} (\bibinfo {year} {2001})},\ \Eprint
  {http://arxiv.org/abs/hep-th/0105057} {arXiv:hep-th/0105057 [hep-th]}
  \BibitemShut {NoStop}%
\bibitem [{\citenamefont {Korepin}(1979)}]{korepin1979direct}%
  \BibitemOpen
  \bibfield  {author} {\bibinfo {author} {\bibfnamefont {V.~E.}\ \bibnamefont
  {Korepin}},\ }\href@noop {} {\bibfield  {journal} {\bibinfo  {journal}
  {Teoreticheskaya i Matematicheskaya Fizika}\ }\textbf {\bibinfo {volume}
  {41}},\ \bibinfo {pages} {169} (\bibinfo {year} {1979})}\BibitemShut
  {NoStop}%
\bibitem [{\citenamefont {Faddeev}\ and\ \citenamefont
  {Korepin}(1978)}]{FADDEEV19781}%
  \BibitemOpen
  \bibfield  {author} {\bibinfo {author} {\bibfnamefont {L.}~\bibnamefont
  {Faddeev}}\ and\ \bibinfo {author} {\bibfnamefont {V.}~\bibnamefont
  {Korepin}},\ }\href {\doibase https://doi.org/10.1016/0370-1573(78)90058-3}
  {\bibfield  {journal} {\bibinfo  {journal} {Physics Reports}\ }\textbf
  {\bibinfo {volume} {42}},\ \bibinfo {pages} {1} (\bibinfo {year}
  {1978})}\BibitemShut {NoStop}%
\bibitem [{\citenamefont {Mandelstam}(1975)}]{PhysRevD.11.3026}%
  \BibitemOpen
  \bibfield  {author} {\bibinfo {author} {\bibfnamefont {S.}~\bibnamefont
  {Mandelstam}},\ }\href {\doibase 10.1103/PhysRevD.11.3026} {\bibfield
  {journal} {\bibinfo  {journal} {Phys. Rev. D}\ }\textbf {\bibinfo {volume}
  {11}},\ \bibinfo {pages} {3026} (\bibinfo {year} {1975})}\BibitemShut
  {NoStop}%
\bibitem [{\citenamefont {Luther}(1976)}]{PhysRevB.14.2153}%
  \BibitemOpen
  \bibfield  {author} {\bibinfo {author} {\bibfnamefont {A.}~\bibnamefont
  {Luther}},\ }\href {\doibase 10.1103/PhysRevB.14.2153} {\bibfield  {journal}
  {\bibinfo  {journal} {Phys. Rev. B}\ }\textbf {\bibinfo {volume} {14}},\
  \bibinfo {pages} {2153} (\bibinfo {year} {1976})}\BibitemShut {NoStop}%
\bibitem [{\citenamefont {Coleman}(1975)}]{PhysRevD.11.2088}%
  \BibitemOpen
  \bibfield  {author} {\bibinfo {author} {\bibfnamefont {S.}~\bibnamefont
  {Coleman}},\ }\href {\doibase 10.1103/PhysRevD.11.2088} {\bibfield  {journal}
  {\bibinfo  {journal} {Phys. Rev. D}\ }\textbf {\bibinfo {volume} {11}},\
  \bibinfo {pages} {2088} (\bibinfo {year} {1975})}\BibitemShut {NoStop}%
\bibitem [{\citenamefont {{Banuls}}\ \emph {et~al.}(2018)\citenamefont
  {{Banuls}}, \citenamefont {{Cichy}}, \citenamefont {{Kao}}, \citenamefont
  {{Lin}}, \citenamefont {{Lin}},\ and\ \citenamefont
  {{Tan}}}]{2018slft.confE.229B}%
  \BibitemOpen
  \bibfield  {author} {\bibinfo {author} {\bibfnamefont {M.~C.}\ \bibnamefont
  {{Banuls}}}, \bibinfo {author} {\bibfnamefont {K.}~\bibnamefont {{Cichy}}},
  \bibinfo {author} {\bibfnamefont {Y.~J.}\ \bibnamefont {{Kao}}}, \bibinfo
  {author} {\bibfnamefont {C.~J.~D.}\ \bibnamefont {{Lin}}}, \bibinfo {author}
  {\bibfnamefont {Y.~P.}\ \bibnamefont {{Lin}}}, \ and\ \bibinfo {author}
  {\bibfnamefont {T.~L.}\ \bibnamefont {{Tan}}},\ }in\ \href@noop {} {\emph
  {\bibinfo {booktitle} {The 36th Annual International Symposium on Lattice
  Field Theory. 22-28 July}}}\ (\bibinfo {year} {2018})\ p.\ \bibinfo {pages}
  {229},\ \Eprint {http://arxiv.org/abs/1810.12038} {arXiv:1810.12038
  [hep-lat]} \BibitemShut {NoStop}%
\bibitem [{\citenamefont {{Ba{\~n}uls}}\ \emph {et~al.}(2019)\citenamefont
  {{Ba{\~n}uls}}, \citenamefont {{Cichy}}, \citenamefont {{Kao}}, \citenamefont
  {{Lin}}, \citenamefont {{Lin}},\ and\ \citenamefont
  {{Tan}}}]{2019PhRvD.100i4504B}%
  \BibitemOpen
  \bibfield  {author} {\bibinfo {author} {\bibfnamefont {M.~C.}\ \bibnamefont
  {{Ba{\~n}uls}}}, \bibinfo {author} {\bibfnamefont {K.}~\bibnamefont
  {{Cichy}}}, \bibinfo {author} {\bibfnamefont {Y.-J.}\ \bibnamefont {{Kao}}},
  \bibinfo {author} {\bibfnamefont {C.~J.~D.}\ \bibnamefont {{Lin}}}, \bibinfo
  {author} {\bibfnamefont {Y.-P.}\ \bibnamefont {{Lin}}}, \ and\ \bibinfo
  {author} {\bibfnamefont {D.~T.~L.}\ \bibnamefont {{Tan}}},\ }\href {\doibase
  10.1103/PhysRevD.100.094504} {\bibfield  {journal} {\bibinfo  {journal}
  {Phys. Rev. D}\ }\textbf {\bibinfo {volume} {100}},\ \bibinfo {eid} {094504}
  (\bibinfo {year} {2019})},\ \Eprint {http://arxiv.org/abs/1908.04536}
  {arXiv:1908.04536 [hep-lat]} \BibitemShut {NoStop}%
\bibitem [{\citenamefont {{Ba{\~n}uls}}\ \emph {et~al.}(2017)\citenamefont
  {{Ba{\~n}uls}}, \citenamefont {{Cichy}}, \citenamefont {{Kao}}, \citenamefont
  {{Lin}}, \citenamefont {{Lin}},\ and\ \citenamefont {{Tao-Lin
  Tan}}}]{2017arXiv171009993B}%
  \BibitemOpen
  \bibfield  {author} {\bibinfo {author} {\bibfnamefont {M.~C.}\ \bibnamefont
  {{Ba{\~n}uls}}}, \bibinfo {author} {\bibfnamefont {K.}~\bibnamefont
  {{Cichy}}}, \bibinfo {author} {\bibfnamefont {Y.-J.}\ \bibnamefont {{Kao}}},
  \bibinfo {author} {\bibfnamefont {C.~J.~D.}\ \bibnamefont {{Lin}}}, \bibinfo
  {author} {\bibfnamefont {Y.-P.}\ \bibnamefont {{Lin}}}, \ and\ \bibinfo
  {author} {\bibfnamefont {D.}~\bibnamefont {{Tao-Lin Tan}}},\ }\href@noop {}
  {\bibfield  {journal} {\bibinfo  {journal} {arXiv e-prints}\ ,\ \bibinfo
  {eid} {arXiv:1710.09993}} (\bibinfo {year} {2017})},\ \Eprint
  {http://arxiv.org/abs/1710.09993} {arXiv:1710.09993 [hep-lat]} \BibitemShut
  {NoStop}%
\bibitem [{\citenamefont {Hotta}(2008{\natexlab{b}})}]{PhysRevD.78.045006}%
  \BibitemOpen
  \bibfield  {author} {\bibinfo {author} {\bibfnamefont {M.}~\bibnamefont
  {Hotta}},\ }\href {\doibase 10.1103/PhysRevD.78.045006} {\bibfield  {journal}
  {\bibinfo  {journal} {Phys. Rev. D}\ }\textbf {\bibinfo {volume} {78}},\
  \bibinfo {pages} {045006} (\bibinfo {year} {2008}{\natexlab{b}})}\BibitemShut
  {NoStop}%
\bibitem [{\citenamefont {{Hotta}}(2011)}]{2011arXiv1101.3954H}%
  \BibitemOpen
  \bibfield  {author} {\bibinfo {author} {\bibfnamefont {M.}~\bibnamefont
  {{Hotta}}},\ }\href@noop {} {\bibfield  {journal} {\bibinfo  {journal} {arXiv
  e-prints}\ ,\ \bibinfo {eid} {arXiv:1101.3954}} (\bibinfo {year} {2011})},\
  \Eprint {http://arxiv.org/abs/1101.3954} {arXiv:1101.3954 [quant-ph]}
  \BibitemShut {NoStop}%
\bibitem [{\citenamefont {Sagawa}\ and\ \citenamefont
  {Ueda}(2008)}]{PhysRevLett.100.080403}%
  \BibitemOpen
  \bibfield  {author} {\bibinfo {author} {\bibfnamefont {T.}~\bibnamefont
  {Sagawa}}\ and\ \bibinfo {author} {\bibfnamefont {M.}~\bibnamefont {Ueda}},\
  }\href {\doibase 10.1103/PhysRevLett.100.080403} {\bibfield  {journal}
  {\bibinfo  {journal} {Phys. Rev. Lett.}\ }\textbf {\bibinfo {volume} {100}},\
  \bibinfo {pages} {080403} (\bibinfo {year} {2008})}\BibitemShut {NoStop}%
\bibitem [{\citenamefont {Lloyd}(1997)}]{PhysRevA.56.3374}%
  \BibitemOpen
  \bibfield  {author} {\bibinfo {author} {\bibfnamefont {S.}~\bibnamefont
  {Lloyd}},\ }\href {\doibase 10.1103/PhysRevA.56.3374} {\bibfield  {journal}
  {\bibinfo  {journal} {Phys. Rev. A}\ }\textbf {\bibinfo {volume} {56}},\
  \bibinfo {pages} {3374} (\bibinfo {year} {1997})}\BibitemShut {NoStop}%
\bibitem [{\citenamefont {Hagen}(1967)}]{hagen1967new}%
  \BibitemOpen
  \bibfield  {author} {\bibinfo {author} {\bibfnamefont {C.}~\bibnamefont
  {Hagen}},\ }\href@noop {} {\bibfield  {journal} {\bibinfo  {journal} {Il
  Nuovo Cimento B (1965-1970)}\ }\textbf {\bibinfo {volume} {51}},\ \bibinfo
  {pages} {169} (\bibinfo {year} {1967})}\BibitemShut {NoStop}%
\bibitem [{\citenamefont {Mueller}\ and\ \citenamefont
  {Trueman}(1971)}]{PhysRevD.4.1635}%
  \BibitemOpen
  \bibfield  {author} {\bibinfo {author} {\bibfnamefont {A.~H.}\ \bibnamefont
  {Mueller}}\ and\ \bibinfo {author} {\bibfnamefont {T.~L.}\ \bibnamefont
  {Trueman}},\ }\href {\doibase 10.1103/PhysRevD.4.1635} {\bibfield  {journal}
  {\bibinfo  {journal} {Phys. Rev. D}\ }\textbf {\bibinfo {volume} {4}},\
  \bibinfo {pages} {1635} (\bibinfo {year} {1971})}\BibitemShut {NoStop}%
\bibitem [{\citenamefont {Gomes}\ and\ \citenamefont
  {Lowenstein}(1973)}]{PhysRevD.7.550}%
  \BibitemOpen
  \bibfield  {author} {\bibinfo {author} {\bibfnamefont {M.}~\bibnamefont
  {Gomes}}\ and\ \bibinfo {author} {\bibfnamefont {J.~H.}\ \bibnamefont
  {Lowenstein}},\ }\href {\doibase 10.1103/PhysRevD.7.550} {\bibfield
  {journal} {\bibinfo  {journal} {Phys. Rev. D}\ }\textbf {\bibinfo {volume}
  {7}},\ \bibinfo {pages} {550} (\bibinfo {year} {1973})}\BibitemShut {NoStop}%
\bibitem [{\citenamefont {Alcaraz}\ and\ \citenamefont
  {Malvezzi}(1995)}]{alcaraz1995critical}%
  \BibitemOpen
  \bibfield  {author} {\bibinfo {author} {\bibfnamefont {F.}~\bibnamefont
  {Alcaraz}}\ and\ \bibinfo {author} {\bibfnamefont {A.}~\bibnamefont
  {Malvezzi}},\ }\href@noop {} {\bibfield  {journal} {\bibinfo  {journal}
  {Journal of Physics A: Mathematical and General}\ }\textbf {\bibinfo {volume}
  {28}},\ \bibinfo {pages} {1521} (\bibinfo {year} {1995})}\BibitemShut
  {NoStop}%
\bibitem [{\citenamefont {Jordan}\ and\ \citenamefont
  {Wigner}(1928)}]{Jordan:1928wi}%
  \BibitemOpen
  \bibfield  {author} {\bibinfo {author} {\bibfnamefont {P.}~\bibnamefont
  {Jordan}}\ and\ \bibinfo {author} {\bibfnamefont {E.~P.}\ \bibnamefont
  {Wigner}},\ }\href {\doibase 10.1007/BF01331938} {\bibfield  {journal}
  {\bibinfo  {journal} {Z. Phys.}\ }\textbf {\bibinfo {volume} {47}},\ \bibinfo
  {pages} {631} (\bibinfo {year} {1928})}\BibitemShut {NoStop}%
%%CITATION = ZEPYA,47,631;%%
\bibitem [{\citenamefont {Kadowaki}\ and\ \citenamefont
  {Nishimori}(1998)}]{PhysRevE.58.5355}%
  \BibitemOpen
  \bibfield  {author} {\bibinfo {author} {\bibfnamefont {T.}~\bibnamefont
  {Kadowaki}}\ and\ \bibinfo {author} {\bibfnamefont {H.}~\bibnamefont
  {Nishimori}},\ }\href {\doibase 10.1103/PhysRevE.58.5355} {\bibfield
  {journal} {\bibinfo  {journal} {Phys. Rev. E}\ }\textbf {\bibinfo {volume}
  {58}},\ \bibinfo {pages} {5355} (\bibinfo {year} {1998})}\BibitemShut
  {NoStop}%
\bibitem [{\citenamefont {{Ikeda}}\ \emph {et~al.}(2019)\citenamefont
  {{Ikeda}}, \citenamefont {{Nakamura}},\ and\ \citenamefont
  {{Humble}}}]{2019NatSR...912837I}%
  \BibitemOpen
  \bibfield  {author} {\bibinfo {author} {\bibfnamefont {K.}~\bibnamefont
  {{Ikeda}}}, \bibinfo {author} {\bibfnamefont {Y.}~\bibnamefont {{Nakamura}}},
  \ and\ \bibinfo {author} {\bibfnamefont {T.~S.}\ \bibnamefont {{Humble}}},\
  }\href {\doibase 10.1038/s41598-019-49172-3} {\bibfield  {journal} {\bibinfo
  {journal} {Scientific Reports}\ }\textbf {\bibinfo {volume} {9}},\ \bibinfo
  {eid} {12837} (\bibinfo {year} {2019})},\ \Eprint
  {http://arxiv.org/abs/1904.12139} {arXiv:1904.12139 [quant-ph]} \BibitemShut
  {NoStop}%
\bibitem [{\citenamefont {Ohzeki}(2020)}]{ohzeki2020breaking}%
  \BibitemOpen
  \bibfield  {author} {\bibinfo {author} {\bibfnamefont {M.}~\bibnamefont
  {Ohzeki}},\ }\href@noop {} {\bibfield  {journal} {\bibinfo  {journal}
  {Scientific reports}\ }\textbf {\bibinfo {volume} {10}},\ \bibinfo {pages}
  {1} (\bibinfo {year} {2020})}\BibitemShut {NoStop}%
\bibitem [{\citenamefont {Kadowaki}\ and\ \citenamefont
  {Nishimori}(2023)}]{kadowaki2023greedy}%
  \BibitemOpen
  \bibfield  {author} {\bibinfo {author} {\bibfnamefont {T.}~\bibnamefont
  {Kadowaki}}\ and\ \bibinfo {author} {\bibfnamefont {H.}~\bibnamefont
  {Nishimori}},\ }\href@noop {} {\bibfield  {journal} {\bibinfo  {journal}
  {Philosophical Transactions of the Royal Society A}\ }\textbf {\bibinfo
  {volume} {381}},\ \bibinfo {pages} {20210416} (\bibinfo {year}
  {2023})}\BibitemShut {NoStop}%
\bibitem [{\citenamefont {{Ikeda}}(2020)}]{2020QuIP...19..331I}%
  \BibitemOpen
  \bibfield  {author} {\bibinfo {author} {\bibfnamefont {K.}~\bibnamefont
  {{Ikeda}}},\ }\href {\doibase 10.1007/s11128-020-02811-5} {\bibfield
  {journal} {\bibinfo  {journal} {Quantum Information Processing}\ }\textbf
  {\bibinfo {volume} {19}},\ \bibinfo {eid} {331} (\bibinfo {year} {2020})},\
  \Eprint {http://arxiv.org/abs/1910.02833} {arXiv:1910.02833 [quant-ph]}
  \BibitemShut {NoStop}%
\bibitem [{\citenamefont {Huh}\ \emph {et~al.}(2021)\citenamefont {Huh},
  \citenamefont {Ikeda}, \citenamefont {Jahnke},\ and\ \citenamefont
  {Kim}}]{PhysRevE.104.024136}%
  \BibitemOpen
  \bibfield  {author} {\bibinfo {author} {\bibfnamefont {K.-B.}\ \bibnamefont
  {Huh}}, \bibinfo {author} {\bibfnamefont {K.}~\bibnamefont {Ikeda}}, \bibinfo
  {author} {\bibfnamefont {V.}~\bibnamefont {Jahnke}}, \ and\ \bibinfo {author}
  {\bibfnamefont {K.-Y.}\ \bibnamefont {Kim}},\ }\href {\doibase
  10.1103/PhysRevE.104.024136} {\bibfield  {journal} {\bibinfo  {journal}
  {Phys. Rev. E}\ }\textbf {\bibinfo {volume} {104}},\ \bibinfo {pages}
  {024136} (\bibinfo {year} {2021})}\BibitemShut {NoStop}%
\end{thebibliography}%

\end{document}